\documentstyle[12pt,pictex,epsfig,amssymb,amsbsy]{article}

\def\G{{G}}
\def\bl{{\beta}}

\title{{ Cusped SYM Wilson loop at 
two loops and beyond}\\[.5cm] }

\author{Yuri Makeenko%
\footnote{Also at the Institute for Advanced Cycling,
Blegdamsvej 19, 2100 Copenhagen \O, Denmark}
\\[.2cm]
{\normalsize
{\it Institute of Theoretical and Experimental Physics}} \\[.1cm]
 \normalsize {\it 117218 Moscow, Russia} \\[.6cm]
Poul Olesen$^*$ \\[.2cm]
 \normalsize{\it The Niels Bohr Institute}\\[.1cm]
  \normalsize {\it Blegdamsvej 17, 2100 Copenhagen \O, Denmark}\\[.6cm]
Gordon W. Semenoff \\[.2cm]
 \normalsize {\it Department of Physics and Astronomy,
University of British Columbia} \\[.1cm]
 \normalsize {\it Vancouver, British Columbia, Canada V6T 1Z1}}

\date{\mbox{}}

\renewcommand{\thefootnote}{\fnsymbol{footnote}}
\newcommand{\newsection}{
\setcounter{equation}{0}
\section}

\def\appendix#1{
  \addtocounter{section}{1}
  \setcounter{equation}{0}
  \renewcommand{\thesection}{\Alph{section}}
  \section*{Appendix \thesection\protect\indent \parbox[t]{11.715cm} {#1}}
  \addcontentsline{toc}{section}{Appendix \thesection\ \ \ #1}
  }

\def\fun#1#2{\lower3.6pt\vbox{\baselineskip0pt\lineskip.9pt
\ialign{$\mathsurround=0pt#1\hfil##\hfil$\crcr#2\crcr\sim\crcr}}}

\newcommand{\bbox}[1]{\boldsymbol{#1}}

\def\e{{\,\rm e}\,}

\def\d{{\rm d}}

\def\i{{\rm i}}

\newcommand{\rf}[1]{(\ref{#1})}
\newcommand{\eq}[1]{Eq.~(\ref{#1})}
\def\be{\begin{equation}}
\def\ee{\end{equation}}
\def\beq{\begin{equation}}
\def\eeq{\end{equation}}
\def\bea{\begin{eqnarray}}
\def\eea{\end{eqnarray}}

\newcommand{\non}{\nonumber \\*}

\newcommand{\ra}{\rightarrow}

\begin{document}          

\begin{titlepage}


\maketitle                 
\vspace*{-5mm}
\begin{abstract}
We calculate the anomalous dimension of the cusped Wilson loop in
${\cal N}=4$ supersymmetric Yang-Mills theory 
to order
$\lambda^2$ ($\lambda=g^2_{YM}N$). We show that the cancellation
between the diagrams with the three-point vertex and the self-energy
insertion to the propagator which occurs for smooth Wilson loops is
not complete for cusped loops, so that an anomaly term remains. This
term contributes to the cusp anomalous dimension. The result agrees
with the anomalous dimensions of twist-two conformal operators with
large spin. We verify the loop equation for cusped loops to order
$\lambda^2$, reproducing the cusp anomalous dimension this way. We
also examine the issue of summing ladder diagrams to all orders. We
find an exact solution of the Bethe-Salpeter equation, summing
light-cone ladder diagrams, and show that
for certain values of parameters it reduces to a Bessel
function. We find that the ladder diagrams cannot reproduce
for large $\lambda$ the $\sqrt{\lambda}$-behavior of the cusp anomalous
dimension expected from the AdS/CFT correspondence.
\end{abstract}

\thispagestyle{empty}
\end{titlepage}

\renewcommand{\thefootnote}{\arabic{footnote}}
\setcounter{footnote}{2}
\setcounter{page}{2}

\newsection{Introduction}

\begin{figure}
\vspace*{3mm}
\centering{
\input{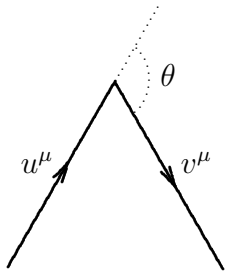}
}
\caption[Cusped Wilson loop]
{Cusped Wilson loop analytically given by \eq{uv}.}
   \label{fi:cusp}
\end{figure}

The cusped Wilson loop has a number of applications to physical
processes. It represents the world trajectory of a heavy quark in
QCD which changes its velocity suddenly at the location of the cusp.
For Euclidean kinematics, or when it is away from the light-cone in
Minkowski space, the cusped Wilson loop is multiplicatively
renormalizable ({for a review see Ref.~\cite{Mak02} and references
therein}). In that case, the ultraviolet divergence is associated
with the bremsstrahlung radiation of soft gluons emitted by the
quark during its sudden change in velocity. The cusp anomalous
dimension $\gamma_{\rm cusp}$ depends on $\theta$, the variable which
represents, as is depicted in Fig.~\ref{fi:cusp}, either the angle
at the cusp in Euclidean space or the change of the rapidity
variable in Minkowski space.

At large $\theta$ the cusp anomalous dimension is proportional to
$\theta$:
\be
\gamma_{\rm cusp} = \frac{\theta}{2} f(g^2_{YM},N)
\,.
\label{largetheta}
\ee
The function $f(g^2_{YM},N)$ can be
calculated perturbatively. In the planar limit, which we shall
discuss exclusively in the following, it is a function of the
't~Hooft coupling $\lambda=g^2_{YM} N$. It is
related~\cite{CD81,KR87,BB88,KM93} to the anomalous dimensions of
twist-two conformal operators~\cite{Bro80} with large spin $J$ in
QCD by 
\be 
\gamma_{J}= f(\lambda) \ln J \,. 
\label{confoper} 
\ee 
In the limit of large $\theta$, the segment of the Wilson loop
approaches the light-cone where the contribution of higher twist
operators are suppressed.

Interest in the anomalous dimensions of the twist-two operators was
inspired by Gubser, Klebanov and Polyakov~\cite{GKP02} who
calculated the leading Regge trajectory of a closed string in type
IIB superstring theory propagating on the background space-time
$AdS_5 \times S^5$. They considered the string rotating on $AdS$
with large angular momentum. According to the AdS/CFT
correspondence~\cite{Mal98a,GKP98,Wit98} (for a review see
Ref.~\cite{AGMOO99}) this trajectory is related to the $\gamma_{J}$
in ${\cal N}=4$ supersymmetric Yang-Mills theory (SYM) and predicts%
\footnote{The $(\sqrt{\lambda})^0$-contribution to this formula,
$-3 \ln2/\pi$, was calculated in Ref.~\cite{FT02}.} \be
f(\lambda)=\frac{\sqrt{\lambda}}{\pi} +{\cal
O}\left((\sqrt{\lambda})^0\right) \label{GKP1} \ee for large
$\lambda$. This result is obtained for large spin $J$, which plays
the role of a semiclassical limit in analogy with the BMN
limit~\cite{BMN02}, and there was argued in Ref.~\cite{GKP02} that
it possesses the features expected for the anomalous dimension in
QCD and may remain to be valid there as well.

Alternatively, the cusp anomalous dimension in ${\cal N}=4$ SYM can
be directly calculated using the
duality~\cite{Mal98b,RY98,BCFM98,DGO99} of the supersymmetric Wilson
loop and an open string in $AdS_5\times S^5$, the ends of which run
along the contour $\{x^\mu(s), \int^s \d s'|\dot x(s')|n^i\}$ 
(where $n^i\in S^5$) at the boundary of $AdS_5\times S^5$.

The relevant Wilson loop~\cite{Mal98b}  
\begin{equation}
W(C) = \frac 1N {\rm
Tr}\,{\bbox P}\,e^{i\int_C ds \dot x^\mu(s) A_\mu(x) + i\int_C
ds |\dot x(s)| n^i\Phi_i(x)}
\label{supersymmetricform}\end{equation} 
contains the
scalar fields $\Phi_i(x)$ of ${\cal N}=4$ SYM as well as the gauge
field. This expression applies to Minkowski space.  In Euclidean
space, the factor of $i$ is absent from the scalar term. When $\dot
x^\mu$ is null, $|\dot x|=0$, as happens when the contour occupies
the light-cone in Minkowski space, the scalar contribution vanishes
and this operator coincides with the usual definition of Wilson loop
in gauge theory.

In the supergravity limit, the string worldsheet coincides with the
minimal surface in $AdS_5\times S^5$ bounded by the loop. Computing
the proper area of this surface determines the asymptotic behavior
of the Wilson loop for large $\lambda$. This approach was first
applied to the cusped Wilson loop, depicted in Fig.~\ref{fi:cusp},
in Ref.~\cite{DGO99}.  It was used in Refs.~\cite{Kru02,Mak03} to
reproduce the strong-coupling result~\rf{GKP1}.

There are a number of circumstances where the strong coupling
asymptotics of Wilson loops can be obtained by summing planar ladder
diagrams. For the case of anti-parallel Wilson lines, for example,
the sum of planar ladders~\cite{ESSZ99} produces the
$\sqrt{\lambda}$ behavior that is found using AdS/CFT~\cite{Mal98b,RY98},
but fails to get the correct coefficient of the quark-antiquark
potential. For the circular Wilson loop, the sum of planar ladder
diagrams can be done explicitly and extrapolated to the strong
coupling limit~\cite{ESZ00} where it is in precise agreement with
the prediction of AdS/CFT. In that case, it has been argued that the
sum of planar ladders obtains the exact result for the Wilson loop
in the 't~Hooft limit. A similar argument can be made for the
correlation functions of chiral primary operators with the circular
Wilson loop which are also thought to be given exactly by the sum of
planar ladder diagrams which can be performed explicitly and agrees
with AdS/CFT~\cite{Semenoff:2001xp}. In addition to providing a test
of AdS/CFT duality, these results make a number of challenging
predictions for IIB superstrings on the $AdS_5 \times S^5$
background~\cite{DG00,Semenoff:2001xp} (for a review see
Ref.~\cite{SZ02}).

A natural question is whether the strong coupling asymptotics
(\ref{GKP1}) for the cusped loop can be obtained from supersymmetric
Yang-Mills perturbation theory. To examine this, we shall begin at
weak coupling by computing the leading perturbative contributions to
the cusped loop to order $\lambda^2$.  In this computation, we
observe that the divergences which lead to the anomalous dimension
of the loop arise from two sources, ladder diagrams and an
incomplete cancellation of divergent diagrams with internal vertices.
The latter is in contrast to smooth loops where, to order
$\lambda^2$, the divergent parts of diagrams with internal vertices
cancel.  In fact, for loops with special geometry, such as the
circle or straight line, their entire contribution cancels. It is
this cancellation which leads to the mild ultraviolet properties of
the SYM Wilson loop (\ref{supersymmetricform}) which were
discussed in Ref.~\cite{DGO99}.  In the case of the cusped loop,
this lack of cancellation already implies that the cusp anomalous
dimension cannot come from ladder diagrams alone, it must obtain
contributions both from diagrams with internal loops and from
diagrams with ladders.

The result for the anomalous dimension of the cusped SYM Wilson loop
to the order $\lambda^2$ agrees with the two-loop anomalous
dimension~\cite{GL80,KL00,KL02,AFK02,KLV03} of twist-two conformal
operators with large spin $J$ in ${\cal N}=4$ SYM, calculated using
regularization via dimensional reduction:
\be
f(\lambda)=\frac{\lambda}{2\pi^2} -\frac{\lambda^2}{96\pi^2}
+\frac{11\lambda^3}{23040 \pi^2}+{\cal O}\left(\lambda^4\right).
\label{order2}
\ee
Here the three-loop term $\sim \lambda^3$ was
obtained in the recent calculation of Ref.~\cite{KLOV}
and reproduced~\cite{Sta05} from the spin-chain Bethe ansatz.
Equation~(\ref{order2}) also agrees with the two-loop
calculation~\cite{KR87,BGK03} of the cusp anomalous dimension of the
non-supersymmetric Wilson loop.

The question that remains asks whether the sum of ladder diagrams
can produce a contribution to the cusp anomalous dimension which
resembles (\ref{GKP1}) at strong coupling. In fact, this was
suggested as the source of the $\sqrt{\lambda}$ strong coupling
behavior in Section~5 of Ref.~\cite{GKP02} in the context of large
spin.  The situation for the cusp could be similar and closely
analogous to that in Ref.~\cite{ESSZ99} for the sum of the ladder
diagrams in the case of antiparallel Wilson lines, there the sum of
ladders exhibits the $\sqrt{\lambda}$-behavior at large $\lambda$
but it is suspected that other diagrams also contribute, and must be
included if the full answer is to match the prediction of AdS/CFT.

In this Paper, we shall examine the issue of summing (rainbow)
ladder diagrams contributing to the cusped loops to all orders. We
shall find an exact solution of the Bethe-Salpeter equation, which
sums light-cone ladder diagrams.  We shall show that for certain
values of parameters it reduces to a Bessel function.

We shall also observe that its asymptotic form indeed contains a
$\sqrt{\lambda}$ term, but it does not contain the leading term in
the rapidity angle $\theta$ in (\ref{largetheta}). This means that
ladders are not the answer at strong coupling, beyond the first few
orders their sum is sub-leading at large $\theta$.  A similar
situation has been observed for the wavy Wilson line \cite{SY04}.
There, to leading order in the waviness, AdS/CFT predicts that the
line has a $\sqrt{\lambda}$ dependence. In that case, by examining
Feynman diagrams, one concludes that the contributions are given
entirely by non-ladder diagrams. In the present case, we could
speculate that the diagrams which contribute are more likely of the
form of the ones
having legs frozen at the location of the cusp, perhaps with several 
lines from internal vertices trapped at the cusp.

In addition, a particular form of the loop equation in ${\cal N}=4$ SYM
was formulated in Ref.~\cite{DGO99}.  
There, it was observed that for the cusped loop
the right-hand side of the loop equation is proportional
to the cusp anomalous dimension calculated 
to one loop order in ${\cal N}=4$ SYM
perturbation theory.  In this Paper, we will re-examine this issue
with an explicit computation to order $\lambda^2$ 
that confirms that for the cusped loop 
the loop equation reproduces the cusp anomalous dimension
to order two loops in SYM perturbation theory.

This Paper is organized as follows. In Sect.~\ref{s:2} we analyze
the two-loop diagrams with internal vertices and show that the
cancellation is not complete in the presence of a cusp. This results
in an appearance of the anomalous boundary term. In Sect.~\ref{s:3}
we calculate the cusp anomalous dimension for Minkowski angles
$\theta$ and find its asymptotic behavior for large $\theta$. The
result agrees with the anomalous dimension of twist-two conformal
operators with large spin. In Sect.~\ref{s:4} we verify the loop
equation for cusped loops to order $\lambda^2$, reproducing the cusp
anomalous dimension this way. In Sect.~\ref{s:5} we find an exact
solution of the Bethe-Salpeter equation,  summing light-cone ladder
diagrams, and show that for certain values of parameters it reduces
to a Bessel function. A conclusion is that the ladder diagrams
cannot reproduce for large $\lambda$ the
$\sqrt{\lambda}$-behavior of the cusp anomalous dimension expected
from the AdS/CFT correspondence. The results and further
perspectives are discussed in Sect.~\ref{s:d}.

\newsection{Graphs with internal vertices\label{s:2}}

\subsection{Kinematics}

Let us parametrize the loop by a function $x_\mu(\tau)$. The cusped
Wilson loop depicted in Fig.~\ref{fi:cusp} is then formed by two
rays:
\begin{equation}
x_\mu (\tau) = \left\{\begin{array}{l}
u_\mu \tau \quad(\tau < 0)  \\
v_\mu \tau  \quad(\tau \geq 0)\,, \\
\end{array} \right.
\label{uv}
\end{equation}
while the cusp is at $\tau=0$. The cusp angle is obviously given by
\begin{eqnarray}
\cos \theta &=& \frac{uv}{\sqrt{u^2}\sqrt{v^2}}
\qquad \hbox{Euclidean space}\,, \nonumber\\*
\cosh \theta &=& \frac{uv}{\sqrt{u^2}\sqrt{v^2}}
\qquad \hbox{Minkowski space} \,.
\label{ch}
\end{eqnarray}

\begin{figure}
\vspace*{3mm}
\centering{
\input{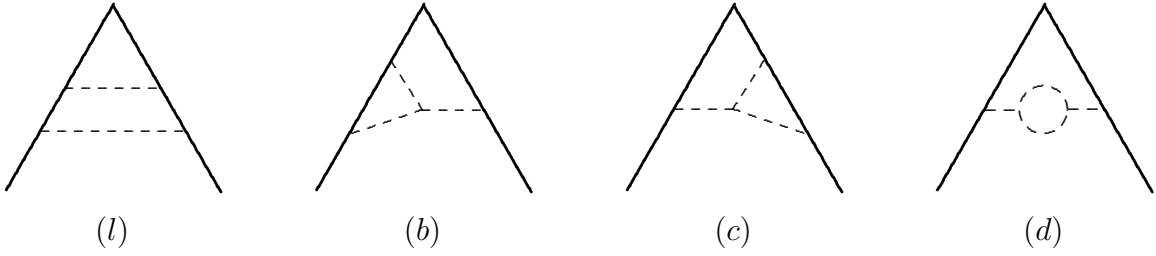}
}
\caption[Two-loop diagrams]
{Two-loop diagrams relevant for the calculation of the anomalous
dimension of the cusped Wilson loop. The dashed lines represent
either the Yang-Mills or scalar propagators.}
   \label{fi:secondo}
\end{figure}

The nontrivial two-loop diagrams that contribute to
the cusped Wilson loop in Fig.~\ref{fi:cusp}
are depicted in Fig.~\ref{fi:secondo}.
The diagram in Fig.~\ref{fi:secondo}$(l)$ is of the type of a ladder with
two rungs. The diagrams in Figs.~\ref{fi:secondo}$(b)$ and $(c)$ involve
the interaction given by
the three-point vertex, while  the diagram in Fig.~\ref{fi:secondo}$(d)$
is of the type of  a self-energy insertion to the propagator.

\subsection{The anomalous surface term}

The analysis of the diagrams  in Figs.~\ref{fi:secondo}$(b)$, $(c)$ and $(d)$
is quite similar to that in the paper
by Erickson, Semenoff and Zarembo~\cite{ESZ00}, where their cancellation
was explicitly shown for straight and circular Wilson loops.

From Eqs.~(13) and (14) of Ref.~\cite{ESZ00}, the sum of all graphs
with one internal three-point vertex in Figs.~\ref{fi:secondo}$(b)$
and $(c)$ is
\begin{eqnarray}
\Sigma_3=-\frac{\lambda^2}{4}\oint d\tau_1 d\tau_2 d\tau_3 \,
\epsilon(\tau_1\tau_2\tau_3)\left( |\dot x^{(1)}| |\dot x^{(3)}|-\dot x^{(1)}
\cdot \dot x^{(3)}\right)
 \dot x^{(2)}\cdot \frac{\partial}{\partial x^{(1)}}
G(x^{(1)}x^{(2)}x^{(3)}) \non
\end{eqnarray}
where $\epsilon(\tau_1\tau_2\tau_3)$ performs antisymmetrization
of $\tau_1$, $\tau_2$ and $\tau_3$ and
the scalar three-point function is
\begin{eqnarray}
\lefteqn{
G(x^{(1)}x^{(2)}x^{(3)})=\int d^{2\omega}w
\Delta( x^{(1)}-w)\Delta( x^{(2)}-w)
\Delta( x^{(3)}-w)}
\non
&=&\frac{\Gamma(2\omega-3)}{64\pi^{2\omega}}
\int  \frac{d\alpha d\beta d\gamma
(\alpha\beta\gamma)^{\omega-2}
\delta(1-\alpha-\beta-\gamma)  }{
\left[ \alpha\beta|x^{(1)}-x^{(2)}|^2+
\beta\gamma|x^{(2)}-x^{(3)}|^2+
\gamma\alpha|x^{(3)}-x^{(1)}|^2
\right]^{2\omega-3}}\,. \non
\end{eqnarray}

Here, we will use a cut cusped trajectory in Euclidean space,
\begin{equation}
x^\mu(\tau)=\left\{ \matrix{ (\tau,0,0,0) & -L<\tau < -\varepsilon
 &{\rm segment~I}\cr
        (\tau\cos\theta, \tau\sin\theta,0,0) &
\varepsilon <\tau < L
&{\rm segment ~II}\cr}
\right.,
\label{parametrized}
\end{equation}
where we must eventually put $L\to\infty$ and $\varepsilon \to 0$.  We
are computing a dimensionless quantity which can only depend on these in
the ratio $\varepsilon/L$.

Because of the presence of $\left( |\dot x^{(1)}| |\dot x^{(3)}|-\dot x^{(1)}
\cdot \dot x^{(3)}\right)$ in the vertex, it vanishes unless $\tau_1$ and
$\tau_3$ are on different segments.

There are four different cases for the integration:
\bea
i)&~&-\frac{\lambda^2}{4}(1-\cos\theta)
\int_{-L}^{-\varepsilon}d\tau_1\int_{-L}^{-\varepsilon}d\tau_2
\int_\varepsilon^Ld\tau_3\epsilon(\tau_1\tau_2)\dot x^{(2)}\cdot
\frac{\partial}{\partial x^{(1)}}\ldots \nonumber \\
ii)&~&-\frac{\lambda^2}{4}(1-\cos\theta)
\int_{-L}^{-\varepsilon}d\tau_1\int_{\varepsilon}^{L}d\tau_2
\int_\varepsilon^Ld\tau_3\epsilon(\tau_2\tau_3)\dot x^{(2)}\cdot
\frac{\partial}{\partial x^{(1)}}\ldots\nonumber \\
iii)&~&-\frac{\lambda^2}{4}(1-\cos\theta)
\int_{\varepsilon}^{L}d\tau_1\int_{-L}^{-\varepsilon}d\tau_2
\int_{-\varepsilon}^{-L}d\tau_3\epsilon(\tau_2\tau_3)\dot x^{(2)}\cdot
\frac{\partial}{\partial x^{(1)}}\ldots\nonumber \\
iv)&~&-\frac{\lambda^2}{4}(1-\cos\theta)
\int_{\varepsilon}^{L}d\tau_1\int_{\varepsilon}^{L}d\tau_2
\int_{-\varepsilon}^{-L}d\tau_3\epsilon(\tau_1\tau_2)\dot x^{(2)}\cdot
\frac{\partial}{\partial x^{(1)}}\ldots
\eea
In case $ii)$, differentiating the expression under the Feynman parameter
integral in $\ldots$ leads to
\bea
\lefteqn{\dot x^{(2)}\cdot
\frac{\partial}{\partial x^{(1)}}
\frac{1 }{
\left[ \alpha\beta|x^{(1)}-x^{(2)}|^2+
\beta\gamma|x^{(2)}-x^{(3)}|^2+
\gamma\alpha|x^{(3)}-x^{(1)}|^2
\right]^{2\omega-3}}} \nonumber \\*
&=&\frac{-(2\omega-3)
\left(
2\alpha\beta(\tau_1\cos\theta-\tau_2)+2\alpha\gamma(\tau_1\cos\theta-\tau_3)
\right)}
{\left[\alpha\beta(\tau_1^2+\tau_2^2-2\tau_1\tau_2\cos\theta)
+\beta\gamma(\tau_2-\tau_3)^2+
\gamma\alpha(\tau_3^2+\tau_1^2-2\tau_3\tau_1\cos\theta)\right]^{2\omega-2} }\,.
\non &&
\eea
We observe that the rest of the integrand is completely antisymmetric under
interchange of $\tau_2,\beta$ and $\tau_3,\gamma$ whereas this term is
symmetric.  Therefore the integral must vanish, $ii)=0$.
By a similar argument, $iii)=0$.

It remains to study $i)$ and $iv)$.  There
$\tau_1$ and $\tau_2$ are on the same segment of the contour and
$\dot x^{(2)}\cdot
\frac{\partial}{\partial x^{(1)}}=\frac{\partial}{\partial \tau_1}$.
Integrating by parts  produces a delta function and a surface term.
Let us first study the term with a delta function.  Using the delta function
to integrate $\tau_2$ results in
\bea
i)&&\frac{\lambda^2}{2}(1-\cos\theta)
\int_{-L}^{-\varepsilon}d\tau_1
\int_\varepsilon^L d\tau_3
\frac{\Gamma(2\omega-3)}{64\pi^{2\omega}}
\int d\alpha d\beta d\gamma \frac{
(\alpha\beta\gamma)^{\omega-2}
\delta(1-\alpha-\beta-\gamma)  }{
\left[  \gamma(1-\gamma)
|x^{(1)}-x^{(3)}|^2
\right]^{2\omega-3}} \non
iv)& &\frac{\lambda^2}{2}(1-\cos\theta)
\int_{\varepsilon}^{L}d\tau_1
\int_{-\varepsilon}^{-L}d\tau_3
\frac{\Gamma(2\omega-3)}{64\pi^{2\omega}}
\int d\alpha d\beta d\gamma \frac{
(\alpha\beta\gamma)^{\omega-2}
\delta(1-\alpha-\beta-\gamma)  }{
\left[  \gamma(1-\gamma)|x^{(1)}-x^{(3)}|^2
\right]^{2\omega-3}}\,. \non &~&
\eea
The result of doing the Feynman parameter integral in
each case is
\begin{equation}\label{i}
i)~~~~\frac{\lambda^2}{2}(1-\cos\theta)
\int_{-L}^{-\varepsilon}d\tau_1
\int_\varepsilon^L d\tau_3
\frac{\Gamma(2\omega-3)}{64\pi^{2\omega}}
\frac{\Gamma^2(\omega-1)}{(2-\omega)\Gamma(2\omega-2)}
 \frac{1 }{
\left[  |x^{(1)}-x^{(3)}|^2
\right]^{2\omega-3}}\,,
\end{equation}
\begin{equation}\label{iv}
iv)~~~~\frac{\lambda^2}{2}(1-\cos\theta)
\int_{\varepsilon}^{L}d\tau_1
\int_{-\varepsilon}^{-L}d\tau_3
\frac{\Gamma(2\omega-3)}{64\pi^{2\omega}}
\frac{\Gamma^2(\omega-1)}{(2-\omega)\Gamma(2\omega-2)}
 \frac{1 }{
\left[  |x^{(1)}-x^{(3)}|^2
\right]^{2\omega-3}}\,.
\end{equation}

The sum of all self-energy contributions to the Wilson loop is given by
(from Eq.~(12) of Ref.~\cite{ESZ00})
\begin{equation}
\Sigma_2=-\frac{\lambda^2\Gamma^2(\omega-1)}
{2^7\pi^{2\omega}(2-\omega)(2\omega-3)}
\oint d\tau_1 d\tau_2 \frac{ |\dot x^{(1)}|
|\dot x^{(2)}|-\dot x^{(1)}\cdot\dot x^{(2)} }
{\left[ (x^{(1)}-x^{(2)})^2\right]^{2\omega-3}}\,.
\end{equation}
It is easy to see that this contribution is canceled by the sum of the
two terms in Eqs.~(\ref{i}) and (\ref{iv}).

Thus, there remains only two contributions, the two surface terms
which are obtained from the integration by parts:
\bea
i)&&-\frac{\lambda^2}{4}(1-\cos\theta)
\int_{-L}^{-\epsilon}d\tau_1\int_{-L}^{-\varepsilon}d\tau_2
\int_\varepsilon^Ld\tau_3\frac{\partial}{\partial\tau_1}\left(
\varepsilon(\tau_1\tau_2) G(x^{(1)},x^{(2)},x^{(3)})\right) \non
&&~~~~~
=-\frac{\lambda^2}{4}(1-\cos\theta)
 \int_{-L}^{-\varepsilon}d\tau_2
\int_\varepsilon^Ld\tau_3
\left(-G(-\varepsilon ,x^{(2)},x^{(3)})
-G(-L, x^{(2)},x^{(3)})\right) \nonumber \\
iv) &&-\frac{\lambda^2}{4}(1-\cos\theta)
\int_{\varepsilon}^{L}d\tau_1\int_{\varepsilon}^{L}d\tau_2
\int_{-\varepsilon}^{-L}d\tau_3\frac{\partial}{\partial\tau_1}\left(
\epsilon(\tau_1\tau_2)G(x^{(1)},x^{(2)},x^{(3)})\right) \non
&& ~~~~~
=-\frac{\lambda^2}{4}(1-\cos\theta)
\int_\varepsilon^Ld\tau_2  \int_{-L}^{-\varepsilon}d\tau_3
\left(-G(L, x^{(2)},x^{(3)})
-G(\varepsilon, x^{(2)},x^{(3)})\right).\non
&& ~~~~~
\eea
These integrals are doubly cutoff, by dimensional regularization and
by $\varepsilon$ and $L$.  Now that we have canceled the singularity in
the bubble diagram, it is safe to remove one of the regulators.  Here we
choose to remove the dimensional regulator to go to the physical dimension
$\omega=2$.  In that case, we have
\begin{eqnarray}
\lefteqn{G(x^{(1)}x^{(2)}x^{(3)})}\non
&=&\frac{1}{64\pi^{4}}
\int d\alpha d\beta d\gamma \frac{
\delta(1-\alpha-\beta-\gamma)  }{
\left[ \alpha\beta(x^{(1)}-x^{(2)})^2+
\beta\gamma(x^{(2)}-x^{(3)})^2+
\gamma\alpha(x^{(3)}-x^{(1)})^2
\right] }.\non &&
\label{defG}
\end{eqnarray}

Then, we find for the two surface terms
\bea
i)&&
=\frac{\lambda^2}{4}(1-\cos\theta)
 \int_{-L/\varepsilon}^{-1}d\tau_2
\int_1^{L/\varepsilon}d\tau_3
\left(G(-1 ,x^{(2)},x^{(3)})
+G(-L/\varepsilon, x^{(2)},x^{(3)})\right) \nonumber \\
iv)&&
=\frac{\lambda^2}{4}(1-\cos\theta)
\int_1^{{L}/{\varepsilon}}d\tau_2
\int_{-{L}/{\epsilon}}^{-1}d\tau_3
\left(G({L}/{\varepsilon}, x^{(2)},x^{(3)})
+G(1, x^{(2)},x^{(3)})\right).
\label{twosurf}
\eea

It is clear that the divergence in these integrals is like log, not log$^2$.
The logarithmically divergent part  is gotten by taking a derivative
$\varepsilon\frac{d}{d\varepsilon}$.
After that, the $\tau$-integrations reduce
to one remaining integration.   Also remaining are the Feynman parameters.
The sum of the logarithmic terms is given by
\be
-2\ln(\varepsilon/L)\int_0^1
\frac{ d\alpha d\beta d\gamma d\tau\,\delta(1-\alpha-\beta-\gamma)}
{\gamma(1-\gamma)\tau^2 +2\cos\theta\beta\gamma\tau+\beta(1-\beta) }\,.
\ee
We can use a symmetry of the integral under $\tau\to 1/\tau$ and interchanging
$\gamma$ and $\beta$ to write it as
\be
-\ln(\varepsilon/L)
\int_0^\infty d\tau\int_0^1\frac{ d\alpha d\beta d\gamma\,
\delta(1-\alpha-\beta-\gamma)}
{\gamma(1-\gamma)\tau^2 +2\cos\theta\beta\gamma\tau+\beta(1-\beta) } \,.
\ee

Finally, we obtain
\be
\gamma_{\rm cusp}^{(b)+(c)+(d)}= -\frac{\lambda^2}{64\pi^4}I
\label{2loopca}
\ee
with
\be
I=
(\cos \theta-1) \int_0^\infty d\tau\int_0^1\frac{ d\alpha d\beta d\gamma\,
\delta(1-\alpha-\beta-\gamma)}
{\gamma(1-\gamma)\tau^2 +2\cos\theta\beta\gamma\tau+\beta(1-\beta) }
\label{defI}
\ee
for the contribution of the two loop
diagrams in Figs.~\ref{fi:secondo}($b$), ($c$) and
($d$) to the cusp anomalous dimension.

It is worth noting once again that Eqs.~\rf{2loopca} and \rf{defI}
result from the surface term which would be missing if there were no
cusp and the loop were smooth. This can be easily seen from
\eq{defI} since $I$ vanishes for $\theta=0$.

\begin{figure}
\vspace*{3mm}
\centering{
\input{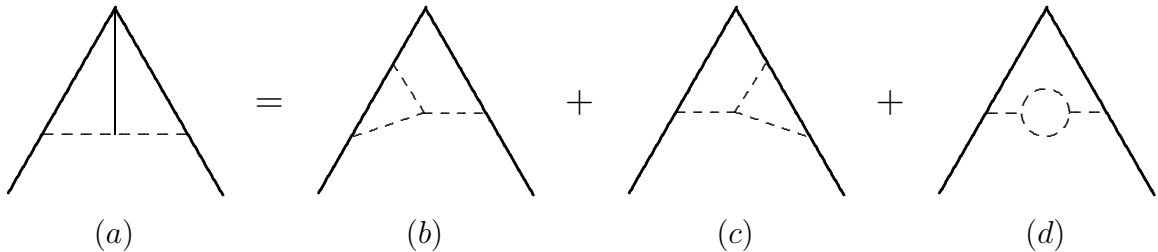}
}
\caption[Anomaly diagram]
{Two-loop anomaly diagram ($a$) which emerges as a surface term in the
sum of the diagrams ($b$), ($c$) and ($d$).
An analytic expression is given
by the sum of $i)$ and $iv)$ in \eq{twosurf}.}
   \label{fi:anom2}
\end{figure}

It is convenient to depict the sum of the two-loop diagrams
($b$), ($c$) and ($d$) as the anomalous diagram in Fig.~\ref{fi:anom2}$(a)$.
All three propagators in Fig.~\ref{fi:anom2}$(a)$ are scalar as it follows
from Eqs.~\rf{twosurf} and \rf{defG}. We shall call this as the
anomalous term.

\newsection{Cusp anomalous dimension at two loops\label{s:3}}

As we have explained in the Introduction, we are interested in the
cusp anomalous dimension for large angle $\theta$ in Minkowski
space. The analytic continuation of \eq{defI} to Minkowski space can
easily be done by passing to the Minkowski variable~\rf{ch} by the
analytic continuation $\theta \ra \i \theta$.  The  result is to
replace  $\cos\theta$ where it appears in  \eq{defI} by
$\cosh\theta$.

\subsection{Arbitrary $\theta$}

Introducing the variables
\be
\tilde \beta = \frac{1}\beta -1,\quad 0< \tilde \beta < \infty\,;
\qquad \beta=\frac1{1+\tilde \beta}
\ee
and
\be
\tilde z=\frac1{\gamma\beta}-\frac1{\gamma}-\frac1{\beta}, \quad
0<\tilde z< \infty\,;\qquad
\gamma=\frac{\tilde\beta}{1+\tilde z+ \tilde \beta}
\ee
we rewrite the integral in \eq{defI} as
\bea
I&=&{(\cosh\theta-1)
\int_0^\infty d\tau\int_0^1 d\beta \int_0^{1-\beta} d\gamma
\frac{1}{\gamma(1-\gamma)\tau^2 +2\gamma\beta\cosh\theta\tau+\beta(1-\beta)}}
\non &=&(\cosh\theta-1)
\int_0^\infty d\tilde\tau\int_1^\infty dz
\frac{1}{z\tilde\tau^2 +2\cosh\theta\tilde\tau+1 }
\int_0^\infty \frac{d\tilde\beta}{(\tilde \beta+1)(\tilde \beta+z)}
\non &=& (\cosh\theta-1)\int_1^\infty
\frac{dz\,{\ln z}}{(z-1)\sqrt{\cosh^2\theta-z}}
\ln \frac{\cosh\theta+\sqrt{\cosh^2\theta-z}}{\sqrt{z}}
\label{uuff}
\eea
with \be z=\tilde z +1\,, \qquad \tilde \tau = \frac{(\tilde
\beta+1)}{(\tilde \beta+z)} \tau\,. \ee The integrand on the
right-hand side of \eq{uuff} is real both of the regions $z<\cosh^2
\theta$ and $ z>\cosh^2 \theta$.  The integral is convergent as $
z\ra 1$ or $ z\ra\infty$.


It is convenient to change the variable $z$ in the integral on the
right-hand side of \eq{uuff} to the angular variable \be \psi=\ln
\frac{\cosh \theta + \sqrt{\cosh^2 \theta -z}}{\sqrt{z}}, \qquad
z=\frac{\cosh^2 \theta}{\cosh^2 \psi}. \ee The variable $\psi$
decreases from $\theta$ to $0$ when $z$ increases from $1$ to
$\cosh^2 \theta$ and then runs along the imaginary axis from $0$ to
$\i \pi/2$ when $z$ further increases from $\cosh^2 \theta$ to
$\infty$. We thus rewrite the integral as \be I= 4\frac{\cosh \theta
-1}{\cosh \theta} \Bigg( \int_0^\theta + \int _0^{\pi/2}\Bigg)\frac{
d\psi \, \psi}{1- \cosh^2 \psi/\cosh^2 \theta} \ln \frac{\cosh
\theta}{\cosh \psi}. \label{I+I} \ee

\subsection{Large $\theta$\label{ss:3}}

The asymptotics at large $\theta$ of the first integral in \eq{I+I}
is governed by
\be
4\int_0^\theta  \frac{d\psi \psi(\theta-\psi)}{1-\e^{2(\psi-\theta)}}
= 4\int_0^\theta  d\psi  \psi(\theta-\psi)
\left({1+\sum_{n=1}^\infty\e^{-2n(\theta-\psi)}}\right).
\label{asymptote}
\ee
Noting that only the values of $(\theta-\psi)\sim 1$ are essential in the
integral in each term of the sum over $n \geq 1$ in \eq{asymptote}
at large $\theta$ with an exponential in $\theta$ accuracy,
we obtain asymptotically
\bea
4\left( \frac{1}6 \theta^3 + \theta \sum_{n=1}^\infty \frac{1}{4n^2}\right)&=&
4\left(\frac{1}6 \theta^3 +\frac{\zeta(2)}4\theta +{\cal O}(1) \right)\non
&=&
 \frac 23 \theta^3+\frac{\pi^2}6\theta
+{\cal O}(1) \,.
\label{resa1}
\eea

A term which is linear in $\theta$  also comes from the second
integral in \eq{I+I}.   For large $\theta$ the latter takes the form
\be 4 \int _0^{\pi/2} d\psi \, \psi \theta = \frac{\pi^2}{2} \theta\,.
\label{resa2} \ee

Summing the asymptotics~\rf{resa1} and \rf{resa2}, we find, the
contributions of the sum of the diagrams in
Figs.~\ref{fi:secondo}$(b)$, $(c)$ and $(d)$: \be \gamma_{\rm
cusp}^{(b)+(c)+(d)}= -\frac{\lambda^2}{96\pi^4} \left( \theta^3
+\pi^2\theta +{\cal O}(1) \right). \label{b+c} \ee

\subsection{Adding the ladder diagram}

The expansion of \eq{b+c} in large $\theta$ begins from a term of
order $\theta^3$.  This term cancels the one coming from the ladder
diagram in Fig.~\ref{fi:secondo}$(l)$. It is convenient to add and
subtract the diagram with crossed propagator lines.  After that, the
sum exponentiates the contribution of the order $\lambda$. The
remaining contribution is proportional to the one calculated in
Ref.~\cite{KR87} and reads as \be \gamma_{\rm cusp}^{(l)}=
\frac{\lambda^2}{128\pi^4} \frac{(\cosh\theta-1)^2}{\sinh^2\theta}
\int_0^\infty \frac{d\sigma}{\sigma} \ln\left( \frac{1+\sigma
\e^\theta}{1+\sigma \e^{-\theta}}  \right) \ln\left( \frac{\sigma
+\e^\theta}{\sigma +\e^{-\theta}}  \right). \label{KRa} \ee This
integral possesses the symmetry $\sigma\ra 1/\sigma$ which makes it
equal twice the integral from 0 to 1.

Introducing the angular variable
\be
\psi=\frac 12 \ln \frac{1+\sigma \e^\theta}{1+\sigma \e^{-\theta}}
\ee
and noting that
\be
\frac {d\sigma}{\sigma}= d\psi \left[\coth \psi + \coth(\theta-\psi) \right],
\ee
we rewrite the integral in \eq{KRa} as
\bea
\lefteqn{\hspace*{-1cm}
\frac{\lambda^2}{16\pi^4} \frac{(\cosh\theta-1)^2}{\sinh^2\theta}
\int_0^\theta d\psi \psi (\theta-\psi)\coth \psi} \non &&=
\frac{\lambda^2}{16\pi^4} \frac{(\cosh\theta-1)^2}{\sinh^2\theta}
\int_0^\theta d\psi \psi (\theta-\psi)
\left(1+2\sum_{n=1}^\infty  \e^{-2n\psi}\right)
\eea
whose asymptotics at large $\theta$ is
\be
\gamma_{\rm cusp}^{(l)}=
\frac{\lambda^2}{16\pi^4}\left(\frac{\theta^3}6 +\frac {\zeta(2)}2 \theta
+ {\cal O}(1) \right)
=\frac{\lambda^2}{96\pi^4} \left(
\theta^3 +\frac {\pi^2} 2 \theta + {\cal O}(1) \right).
\label{lc}
\ee

Summing up the contributions of all of the four diagrams in
Fig.~\ref{fi:secondo}, we finally obtain  \be \gamma_{\rm cusp}=
\frac \theta 2 \left(\frac{\lambda}{2\pi^2}
-\frac{\lambda^2}{96\pi^2}\right) +{\cal O}(\theta^0) \label{cusp2}
\ee for the cusp anomalous dimension to order $\lambda^2$.  This
expression agrees with the result in \eq{order2} which is obtained
by different methods.

\newsection{Cusped loop equation to order $\lambda^2$\label{s:4}}

The dynamics of Wilson loops in Yang-Mills theory is
governed by the loop equation~\cite{MM79}
(for details see Ref.~\cite{Mak02}, Chapter~12).
An extension of the loop equation
to ${\cal N}=4$ SYM was proposed by
Drukker, Gross and Ooguri~\cite{DGO99}.
It deals with supersymmetric loops
${\bbox C}=\{x_\mu(\sigma), Y_i(\sigma);
\zeta(\sigma)\}$, where $\zeta(\sigma)$ denotes the Grassmann odd component.

For cusped Wilson loops this loop equation reads as \bea {\bbox
\Delta} \ln W({\bbox C})\Big|_{{\bbox C}=\Gamma} &= &\lambda \int \d
\sigma_1 \int \d \sigma_2 \, \left( \dot x_\mu(\sigma_1)\dot
x_\mu(\sigma_2)- |\dot x_\mu(\sigma_1)||\dot x_\mu(\sigma_2)|
\right) \non &&~~~~~\times
\delta^{(4)}{(x_1-x_2)}\frac{W(\Gamma_{x_1
x_2})W(\Gamma_{x_2x_1})}{W(\Gamma)} \,, \label{cle} \eea where
\be {\bbox \Delta} = \lim_{\eta\to0} \int \d s
\int_{s-\eta}^{s+\eta} \d s' \left( \frac{\delta^2}{\delta
x^\mu(s')\delta x_\mu(s)}+ \frac{\delta^2}{\delta Y^i(s')\delta
Y_i(s)}+ \frac{\delta^2}{\delta \zeta(s')\delta \bar\zeta(s)}\right)
\ee 
is the supersymmetric extension of the loop-space Laplacian 
and we put $\dot Y^2=\dot x^2$, $\zeta=0$ after acting by
${\bbox \Delta} $. The coefficient on the right-hand side of
\eq{cle} accounts for the fact that the Wilson loops are in the
adjoint representation.

Equation~\rf{cle} is in the spirit of the general form of the loop
equation applicable for scalar theory~\cite{Mak88}. For later
convenience the operator on the left-hand side acts on $\ln W$
rather than on $W$ itself and, correspondingly, the right-hand side
is divided by $W(\Gamma)$. This is possible because ${\bbox \Delta}$
is an operator of first order (obeys the Leibnitz rule). A different
but equivalent operator was used in Ref.~\cite{FKKT97} for deriving
loop equations in the IIB matrix model.

It was argued in Ref.~\cite{SY04} that an infinite straight Wilson
line is a solution of the ${\cal N}=4$ SYM loop equation. In
Ref.~\cite{DGO99} it was shown to
order $\lambda$ that the right-hand side of the cusped
loop equation is proportional to the cusped anomalous dimension, 
when the presence of $W(\Gamma)$ 
in the denominator was not essential to order $\lambda$
since $W=1+{\cal O}(\lambda)$. We shall demonstrate in this Section
that this also works for the cusped loop equation~\rf{cle} to the
order $\lambda^2$.

\subsection{The set up}

The delta-function on the right-hand side of \eq{cle} has to be
regularized consistently with the UV regularization of the
propagator, {\em e.g.} \be D_a (x) = \frac{1}{4\pi^2 (x^2+a^2)}\,,
\qquad \delta^{(4)}_a{(x)}= \frac{2 a^2}{\pi^2 (x^2+a^2)^3} \,,
\label{Ddreg} \ee where $a$ is a UV cutoff.

In contrast to the usual Yang-Mills loop equation~\cite{MM79}, the
term of the order $L/a^3$ on the right-hand side of \eq{cle} coming
from the Yang-Mills field is canceled by scalars. For smooth loops
the right-hand side of \eq{cle} will be of the order $(La)^{-1}$
where $L$ is a typical size of the loop. Analogously, the operator
${\bbox \Delta}$ on the left-hand side will produce a term of the same
order. The situation changes for cusped Wilson loops, when the
right-hand side is estimated to be of the order $a^{-2}$ which is
much larger. So the loop equation has specific features for cusped
loops. We are thus going to verify \eq{cle} at the order $a^{-2}$.

An analysis of diagrams with the regularized propagator~\rf{Ddreg}
suggests that the action of the loop-space Laplacian $\Delta$
on each diagram
can be replaced for the cusped Wilson loops  to the order $a^{-2}$
by a differentiation with respect to $a$:
\be
\Delta \ln W(C)\Big|_{C=\Gamma} = 2 \left( \frac 1a \frac{\d}{\d a}-
\frac{\d^2}{\d a^2}  \right)\ln W(\Gamma)+{\cal O}\left( a^{-1} \right)\,.
\label{Ddaa}
\ee
This prescription follows from the formula
\be
-2 \left( \frac 1a \frac{\d}{\d a}-
\frac{\d^2}{\d a^2}  \right)D_a(x)=\delta^{(4)}_a (x)
\ee
and it can be shown for the diagrams of
the orders $\lambda$ and $\lambda^2$.

We therefore conjecture that \be {\bbox \Delta} \ln W({\bbox
C})\Big|_{{\bbox C}=\Gamma} = \frac{2}{a^2} \gamma_{\rm
cusp}\left(\theta,\lambda\right) +{\cal O}\left( a^{-1} \right)
\label{master} \ee so that \eq{cle} reduces to \bea \frac{2}{a^2}
\gamma_{\rm cusp}\left(\theta,\lambda\right) &= &\lambda \int \d
\sigma_1 \int \d \sigma_2 \, \left( \dot x_\mu(\sigma_1)\dot
x_\mu(\sigma_2)- |\dot x_\mu(\sigma_1)||\dot x_\mu(\sigma_2)|
\right) \non &&~~~~~\times
\delta^{(4)}_a(x_1-x_2)\frac{W(\Gamma_{x_1
x_2})W(\Gamma_{x_2x_1})}{W(\Gamma)} \label{mcle} \eea for the cusped
Wilson loops to the order $a^{-2}$. We shall verify \eq{mcle} to the
order $\lambda^2$ by calculating the right-hand side.

To the order $\lambda$ the ratio of the $W$'s on the right-hand side
of \eq{cle} equals 1, and we have \bea && \lambda \int \d \sigma_1
\int \d \sigma_2 \, \left( \dot x_\mu(\sigma_1)\dot x_\mu(\sigma_2)-
|\dot x_\mu(\sigma_1)||\dot x_\mu(\sigma_2)| \right)
\delta^{(4)}_a{(x_1-x_2)}\non &&~~~= 2\lambda (\cosh \theta
-1)\int_0^\infty \d S \int_0^\infty \d T \, \frac{2 a^2}{\pi^2
(S^2+2 ST \cosh \theta +T^2 +a^2)^3} \,. \label{leo1} \eea Changing
the variables for the radial variable $r$ and the angular variable
$\nu$: \be S=\frac{\sqrt{r}}{\sqrt{\nu^2 +2 \nu \cosh \theta+1}}\,,
\qquad T = \frac{\nu\sqrt{r}}{\sqrt{\nu^2 +2 \nu \cosh \theta+1}}\,,
\ee we rewrite \eq{leo1} as \bea &&\frac2{\pi^2} \lambda (\cosh
\theta -1) \int_0^\infty \d r \,\frac{a^2}{(r+a^2)^3} \int_0^\infty
\d \nu\, \frac{1}{\nu^2 +2 \nu \cosh \theta+1}\non
&&~~~=\frac1{2\pi^2 a^2} \lambda \frac{(\cosh \theta
-1)}{\sinh\theta}\theta \eea which agrees with \cite{DGO99} and to
the order $\lambda$ reproduces~\rf{order2} through \eq{mcle}.

\subsection{Ladder contribution}

To the order $\lambda^2$ we substitute on the  right-hand side of
\eq{cle} the ratio of the $W$'s from the previous order $\lambda$,
so that only (minus) the diagrams with crossed lines (one propagator
line and one line representing $\delta^{(4)}_a(x_1-x_2)$) are left
after the cancellation. We obtain for the right-hand side \bea
&&\frac 12 \lambda^2(\cosh \theta -1)^2 \int_0^\infty \d S
\int_0^\infty \d T \left(\int_0^S \d s \int_T^\infty \d t +
\int_S^\infty \d s \int_0^T \d t \right)\non &&~~~\times \frac{2
a^2}{\pi^2 (S^2+2 ST \cosh \theta +T^2 +a^2)^3} \frac{1}{4\pi^2
(s^2+2 st \cosh \theta +t^2 +a^2)} \eea which gives after the
separation of radial and angular variables \bea
&&\frac{\lambda^2}{2\pi^4} (\cosh \theta -1)^2 \int_0^\infty \d r
\,\frac{a^2}{(r+a^2)^3} \int_0^\infty \d \nu\, \frac{1}{\nu^2 +2 \nu
\cosh \theta+1}\non &&~~~\times\int_0^1 \d\sigma \int_\nu^\infty \d
\tau \,
\frac{1}{\sigma^2 +2 \sigma \tau \cosh \theta+\tau^2} \nonumber \\
&&=\frac{\lambda^2}{64\pi^4 a^2}
\frac{(\cosh\theta-1)^2}{\sinh^2\theta} \int_0^\infty
\frac{\d\tilde\nu}{\tilde\nu} \ln\left( \frac{1+\tilde\nu
\e^\theta}{1+ \tilde\nu\e^{-\theta}}  \right) \ln\left(
\frac{\tilde\nu +\e^\theta}{\tilde\nu +\e^{-\theta}}  \right),
\label{thesame} \eea where we substituted \be s=\sigma S\,, \qquad
t=\tau S\,, \qquad \nu =\sigma \tilde\nu\,. \ee The expression on
the right-hand side of \eq{thesame} is the same as in \eq{KRa}.

\subsection{Anomaly contribution}

In order to find the contribution from the anomaly, we
need to be more careful with the regularization.
The regularization~\rf{Ddreg} violate, in general, the ${\cal N}=4$
supersymmetry. We shall instead regularize by the standard dimensional
reduction, which preserve the supersymmetry. Otherwise we cannot
expect the cancellation described in Sect.~\ref{s:2}.

\begin{figure}
\vspace*{3mm}
\centering{
\input{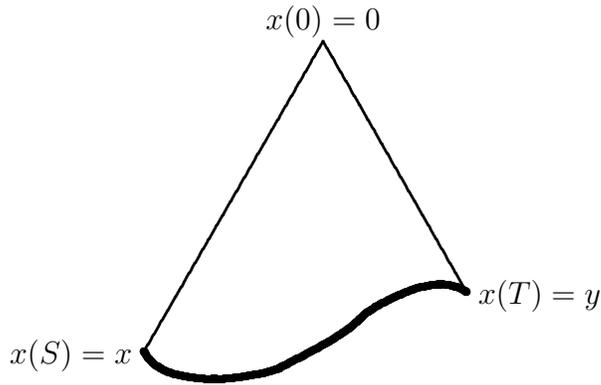}
} \caption[Regularizing path] {Typical path $z(t)$ (represented by
the bold line) connecting $x=z(0)$ and $y=z(\tau)$ in the
regularization of the delta-function on the right-hand side of the
loop equation by \eq{pidreg}. A typical length of the path is $\sim
a$.}
   \label{fi:regpath}
\end{figure}

However, the dimensional regularization does not properly regularize
the delta-func\-tion on the right-hand side of \eq{cle}. We shall
introduce additionally the smearing of the delta-function by \be
\delta^{(d)}(x-y)\longrightarrow \int\limits_{\rm reg.} \d\tau\,
\frac{\partial}{\partial \tau} \frac{1}{(2\pi\tau)^{d/2}}
\e^{-(x-y)^2/2\tau} \label{dreg} \ee
or in the form preserving gauge invariance 
by the path integral of the path-ordered
exponential:
\be
\delta^{(d)}(x-y)\longrightarrow
\int\limits_{\rm reg.} \d\tau\, \frac{\partial}{\partial \tau}
\int\limits_{z(0)=x \atop z(\tau)=y}
{\cal D}z(t)\e^{-\int_0^\tau \d t \,\dot z^2(t)/2}
\,
{\bbox P}\e^{\i \int^\tau_0 \d t\, \dot z^\mu A_\mu(z)}\,.
\label{pidreg} \ee A typical path $z(t)$ connecting $x=z(0)$ and
$y=z(\tau)$ in the regularization of the delta-function on the
right-hand side of the loop equation~\rf{cle} by the path
integral~\rf{pidreg} is depicted in Fig.~\ref{fi:regpath} by the
bold line. A typical length of this path is $\sim a$.

In Eqs.~\rf{dreg} and \rf{pidreg} the integral over the proper time $\tau$
is cut at $\tau\sim a^2$ with $a$ being the UV cutoff.
The regularization ~\rf{Ddreg} is associated with
\be
\int\limits_{\rm reg.} ~\cdots= \int_0^\infty \e^{-a^2/2\tau}~\cdots
\label{reg1}
\ee
while the Schwinger proper-time regularization is given by
\be
\int\limits_{\rm reg.} ~\cdots= \int_{a^2}^\infty ~\cdots\,.
\label{reg2}
\ee

Introducing a UV cutoff $a$ on the right-hand side of the loop
equation inevitably results in a UV regularization of propagators at
distances $\sim a$. To preserve SUSY we choose $a$ to be much
smaller than the UV cutoff, provided by the dimensional
regularization.

\begin{figure}
\vspace*{3mm}
\centering{
\input{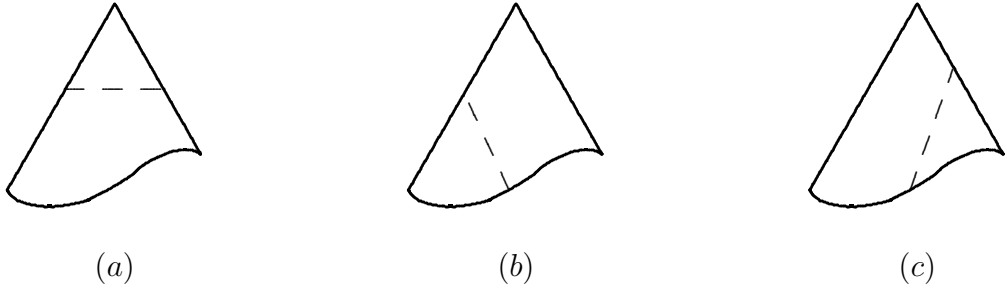}
}
\caption[Diagrams of order $\lambda$]
{Diagrams of order $\lambda$ for closed cusped Wilson loop.
The diagram $(a)$ is the usual one. For the diagrams $(b)$ and $(c)$
one end of the propagator line ends at the regularizing path.}
   \label{fi:dia1}
\end{figure}

In order to calculate the right-hand side of the loop equation to
the order $\lambda^2$, we need the Wilson loop average to the order
$\lambda$. A nonvanishing contribution comes from the diagrams
depicted in Figs.~\ref{fi:dia1}$(a)$, $(b)$ and $(c)$.
The diagram in Fig.~\ref{fi:dia1}$(a)$ is the usual one, while for
the diagrams $(b)$ and $(c)$ one end of the propagator line ends at
the path regularizing the delta-function. The length of this path
from $x$ to $y$ is $\sim\sqrt{\tau}$, which we do not consider to be
small because the contribution of this diagram to the
right-hand side of \eq{cle} will also be $\sim 1/a^2$ for $\tau \sim
a^2$.

\begin{figure}
\vspace*{3mm}
\centering{
\input{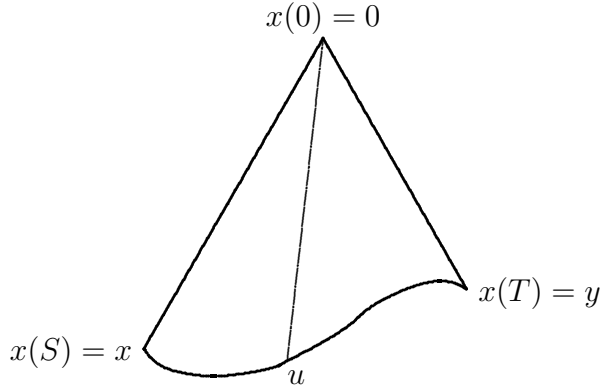}
} \caption[Anomaly diagram] {Two-loop anomaly diagram on the
right-hand side of the loop equation~\rf{cle}. An analytic
expression is given by \eq{to3}.}
   \label{fi:path2a}
\end{figure}

The diagram in Fig.~\ref{fi:dia1}$(a)$ results in the ladder
diagram which we have already considered in the previous subsection.
The diagrams in Figs.~\ref{fi:dia1}$(b)$  and $(c)$ result
in the anomaly diagram of type depicted in Fig.~\ref{fi:path2a}.

The latter statement can be proved by virtue of the useful formula
\bea
\lefteqn{\hspace*{-1cm}\int\limits_{z(0)=x \atop z(\tau)=y}
{\cal D}z(t)\e^{-\int_0^\tau \d t \,\dot z^2(t)/2}
{\int^y_x \d z^\mu \delta^{(d)}(z-u)} = \frac 12
\int_0^\infty \d \tau_1 \int_0^\infty \d \tau_2\,
\delta \left(\tau -\tau_1-\tau_2\right)\hspace*{3cm}} \non
&&\hspace*{3cm}\times \frac{1}{(2\pi \tau_1)^{d/2}}\e^{-(x-u)^2 /2 \tau_1}
\frac{\stackrel{\leftrightarrow}{\partial} }{\partial u_\mu}
\frac{1}{(2\pi \tau_2)^{d/2}}\e^{-(y-u)^2 /2 \tau_2}
\eea
which can be derived using the technique of Ref.~\cite{MM81}%
\footnote{See also Ref.~\cite{Mak02}, pp.~32--33.}.
The geometry is shown in Fig.~\ref{fi:path2a}.
The anomaly diagram in Fig.~\ref{fi:path2a} then appears
by the same token as in Sect.~\ref{s:2}.
We believe that it also works for the regularization
\be
\delta^{(d)}(x-y)\longrightarrow
\int\limits_{\rm reg.} \d\tau\, \frac{\partial}{\partial \tau}
\int\limits_{z(0)=x \atop z(\tau)=y}
{\cal D}z(t)\e^{-\int_0^\tau \d t \,\dot z^2(t)/2}
\,
{\bbox P}\e^{\i \int^\tau_0 \d t \left( \dot z^\mu A_\mu(z)
+ |\dot z| n^i \Phi_i(z)\right)}
\label{pidregss}
\ee
which is both gauge-invariant and supersymmetric.

The contribution of the anomaly diagram in Fig.~\ref{fi:path2a} to
the right-hand side of \eq{cle} is given by \bea &&\hspace*{-.5cm}
\frac {\lambda^2}4 \int_0^\infty \d S\int_0^\infty \d T \,
\left(\dot x_\mu(S) \dot x_\mu(T) - |\dot x(S)\!|| \dot
x(T)\!|\right)\, \int\limits_{\rm reg.} \d\tau\,
\frac{\partial}{\partial \tau} \int_0^\infty \d \tau_1 \int_0^\infty
\d \tau_2\int_0^\infty \d \tau_3 \non &&\hspace*{-.5cm}~~~~\times
\delta \left(\tau -\tau_1-\tau_2\right) \int \d^d
u\,\frac{\e^{-(x-u)^2 /2 \tau_1}}{(2\pi \tau_1)^{d/2}}
\frac{\e^{-(y-u)^2 /2 \tau_2}}{(2\pi \tau_2)^{d/2}}
\frac{\e^{-u^2 /2 \tau_3}}{(2\pi \tau_3)^{d/2}}\non
&&\hspace*{-.5cm}~~=-\frac {\lambda^2}{4 (2\pi)^d} \left(\cosh
\theta -1\right) \Gamma(2\omega-1)
\int_0^\infty \d S \int_0^\infty \d T \non
&&\hspace*{-.5cm}~~~~\times
\int_0^1
\frac {\d \alpha \,\d \beta \,\d \gamma \, (\alpha \beta \gamma)^{\omega-2}
\delta(1-\alpha-\beta -\gamma)\beta^2\gamma^2/(\beta+\gamma)^2
}{[\beta(1-\beta)S^2 + 2 \cosh \theta \beta \gamma ST
+ \gamma (1-\gamma)T^2  + a^2 \beta^2\gamma^2/(\beta+\gamma)^2
]^{2\omega-1}}\,,\non &&
\label{to3}
\eea
where $x(S)=x$, $x(T)=y$ and $x(0)=0$ and we have used
the regularization~\rf{reg1}.

The integral on the right-hand side of \eq{to3} can be treated
similarly to \eq{uuff}. Introducing the radial variable $r$ and the
angular variable $\nu$ by \bea T&= &\frac {\sqrt{r}}
{\sqrt{\gamma(1-\gamma)\nu^2 +2 \beta \gamma \nu \cosh \theta  +
\beta (1-\beta)}}\,, \non S &=&\frac {\nu \sqrt{r}}
{\sqrt{\gamma(1-\gamma)\nu^2 +2 \beta \gamma \nu \cosh \theta  +
\beta (1-\beta)}}\,, \eea we find in $d=4$ \bea &&\protect{{\rm
Eq.~\rf{to3}}}= -\frac {\lambda^2}{64\pi^2}(\cosh \theta-1) \int
_0^\infty \d \nu \non && ~~\times \int_0^1 \frac{\d \alpha\, \d
\beta\, \d \gamma\, \delta(1-\alpha-\beta -\gamma) \beta^2 \gamma^2
/(\beta+\gamma)^2} {\gamma(1-\gamma)\nu^2 +2 \beta \gamma \nu \cosh
\theta  + \beta (1-\beta) } \int _0^\infty \d r
\frac{a^2}{(r+a^2\beta^2 \gamma^2 /(\beta+\gamma)^2)^3}~~~~~~\non
&&~~~~=-\frac {\lambda^2}{32 \pi^2 a^2}(\cosh \theta-1)\int_1^\infty
\frac{\d z\, \ln z}{(z-1) \sqrt{\cosh^2\theta-z}} \ln
\frac{\cosh\theta+\sqrt{\cosh^2\theta-z}}{\sqrt{z}} \,.\non
&&~~~~ \eea This is
the same integral as in \eq{uuff}

\newsection{Ladder diagrams\label{s:5}}

Summing the ladder diagrams is the simplest way to go beyond perturbation
theory. They belong to the class of rainbow diagrams whose important
role in the AdS/CFT correspondence is already recognized~\cite{ESZ00}.
For the cusped Wilson loop depicted in Fig.~\ref{fi:cusp}, the ladder
and rainbow diagrams are essentially the same because the rainbow
diagrams vanish if a propagator has the ends at the same ray of
the loop.

Our idea will be to sum up the ladder diagrams for the cusped SYM Wilson loop,
and to investigate  whether or not they can reproduce
the $\sqrt{\lambda}$-behavior
of the cusp anomalous dimension at large $\lambda$.
This is motivated by the results of Ref.~\cite{ESSZ99}
for the sum of the ladder diagrams in the case of antiparallel Wilson lines.

\subsection{The ladder equation}

The Bethe-Salpeter equation for the sum of the ladder diagrams  is
similar to that of Ref.~\cite{ESSZ99}.

Let the parameter $\tau$ in \eq{uv} be: $\tau=t$ for $\tau>0$ and $\tau=-s$
for $\tau<0$. Let
$s\in [a,S]$ and $t\in[b,T]$, i.e.\ $a$ and $b$ are lower limits for the
integration over $s$ and $t$, respectively. Correspondingly, $S$ and $T$
are the upper limits. We denote the sum of such defined ladder graphs as
$\G\left(S,T;a,b\right)$. It would play the role of
a kernel in an exact Bethe-Salpeter equation, summing all
diagrams not only ladders. In particular, it appears in the
cusped loop equation~\rf{cle}.

It we pick up the first (closest to the cusp) rung of the ladder,
we obtain the equation
\begin{equation}
\G\left(S,T;a,b\right)=1-\frac{\lambda}{4 \pi^2} (\cosh \theta-1)
\int_a^S ds \int_b^{ T} dt\,
\frac{\G\left(S,T;s,t\right)}{s^2 + 2 st \cosh \theta + t^2}\,.
\label{tladder1}
\end{equation}
It we alternatively pick up the last (farthest from the cusp) rung
of the ladder,  we get the equation
\begin{equation}
\G\left(S,T;a,b\right)=1-
\frac{\lambda}{4 \pi^2} (\cosh \theta-1)
\int_a^S ds \int_b^{ T} dt\,
\frac{\G\left(s,t;a,b\right)}{s^2 + 2 st\cosh \theta + t^2} \,.
\label{tladder2}
\end{equation}

In order to find an iterative solution of the ladder equation,
it is convenient first to account for exponentiation by introducing
\begin{equation}
F(S,T;a,b)=- \ln \G (S,T;a,b)\,.
\label{FG}
\end{equation}
Then the ladder equation (\ref{tladder2}) takes the form (this can
be shown by converting the equation for $\G$ to a differential
equation, then substituting $\G=e^{-F}$, then re-integrating)
\begin{eqnarray}
F(S,T;a,b)&=& \frac{\lambda}{4\pi^2}(\cosh \theta-1) \int_a^S ds \int_b^T dt \,
\frac{1}{s^2 + 2 st \cosh \theta + t^2}  \non &&
+\int_a^S ds \int_b^T dt \,
\frac{\partial F(s,t;a,b)}{\partial s}
\frac{\partial F(s,t;a,b)}{\partial t} \,.
\label{ladexpo}
\end{eqnarray}

The order $\lambda$ is given by the first term on the
right-hand side and we obtain
\begin{eqnarray}
F_1(S,T;a,b)&=&\frac{\lambda}{4\pi^2} \frac{\cosh \theta-1}{2\sinh \theta}
\left( L_2(-\frac T{S} e^\theta)-L_2(-\frac T{ S} e^{-\theta})
-L_2(-\frac T{a} e^{\theta})+
L_2(-\frac T{ a} e^{-\theta}) \right.\nonumber \\*
&&\hspace*{2cm} \left.-L_2(-\frac b{S}
e^{\theta})+L_2(-\frac b{S} e^{-\theta})
+L_2(-\frac b{a} e^{\theta})-
L_2(-\frac b{a} e^{-\theta}) \right).
\non &&
\label{tF1}
\end{eqnarray}
Here $L_2$ is Euler's dilogarithm
\begin{equation}
L_2(z)= \sum_{n=1}^\infty \frac{z^n}{n^2}=-\int_0^z \frac{d x}{x}\,
\ln{\left(1-x\right)}
\end{equation}
which obeys the relation%
\footnote{Bateman manuscript on Higher Transcendental Functions, Sect.~1.11.1.}
\begin{equation}
L_2\left(-e^\Omega\right)+L_2\left(-e^{-\Omega}\right)=
-\frac 12 \ln^2 \Omega -\frac{\pi^2}6 \,.
\label{Eul}
\end{equation}
Using (\ref{Eul}), it can be shown that \eq{tF1} possesses
a proper symmetry under interchange
of $S$, $a$ and $T$, $b$.

When $S=\infty$ four terms on the right-hand side of \eq{tF1} vanish
and we find
\begin{equation}
F_1(S=\infty,T;a,b\sim a)
\stackrel{T\gg a}=\frac{\lambda}{4\pi^2} \frac{\cosh\theta-1}{\sinh \theta}\,
\theta \ln \frac Ta
\end{equation}
for $\ln (T/a)\gg1$ and $a\sim b$.

The order $\lambda^2$ can be obtained by inserting $F_1$ into
the second term on the right-hand side of Eq.~(\ref{ladexpo}). This
gives
\begin{eqnarray}
F_2(S,T;a,b) &= &
\frac{\lambda^2}{16\pi^4} (\cosh\theta-1)^2
\int_a^S ds_1 \int_b^T dt_2 \non & & \times
\int_b^{t_2} \frac{dt_1}{(s_1^2+2\cosh\theta s_1t_1+t_1^2)}
\int_a^{s_1} \frac{ds_2}{(s_2^2+2\cosh\theta s_2t_2+t_2^2)}
\non &&
\label{tcrossedladders}
\end{eqnarray}
which is nothing but the diagram with crossed ladders.
This expression can be easily integrated twice.

By the differentiation of the result
with respect to $a$ and $b$, we get
\begin{eqnarray}
\lefteqn{\hspace*{-.3cm}- a \frac d{da} F_2(S,T;a,b)\Big|_{S=T=\infty} }
\non &=&
\frac{\lambda^2}{16\pi^4}  \left(
\frac{\cosh\theta-1}{2\sinh \theta}\right)^2
\int_b^\infty \frac {dt}t \ln \left(\frac{a+t\,e^\theta}{a+t\,e^{-\theta}}
\right)\ln \left(\frac{t+a\,e^\theta}{t+a\,e^{-\theta}}
\right), \nonumber \\
\lefteqn{\hspace*{-.3cm}- b \frac d{db} F_2(S,T;a,b)\Big|_{S=T=\infty} }\non
&=& \frac{\lambda^2}{16\pi^4}
  \left(
\frac{\cosh\theta-1}{2\sinh \theta}\right)^2
\int_a^\infty \frac {ds}s \ln \left(\frac{b+s\,e^\theta}{b+s\,e^{-\theta}}
\right)\ln \left(\frac{s+b\,e^\theta}{s+b\,e^{-\theta}}
\right) \non &&
\label{Pft2}
\end{eqnarray}
so that
\begin{eqnarray}
\lefteqn{- \left(a \frac d{da}+  b \frac d{db} \right)
F_2(S,T;a,b)\Big|_{S=T=\infty}}\non&& =
\frac{\lambda^2}{16\pi^4}  \left(
\frac{\cosh\theta-1}{2\sinh \theta}\right)^2
\int_0^\infty \frac {d z}z \ln \left(\frac{1+z\,e^\theta}{1+z\,e^{-\theta}}
\right)\ln \left(\frac{z+e^\theta}{z+e^{-\theta}}
\right).
\end{eqnarray}
This expression is universal (does not depend on the ratio $b/a$)
and reproduces the result \rf{KRa} of Ref.~\cite{KR87}
for the contribution of the ladders to the cusp anomalous dimension.
We have therefore justified the procedure of Sect.~\ref{s:2}
to use $a\sim\varepsilon$ and $b\sim\varepsilon$ as an ultraviolet cutoff.

\subsection{Light-cone limit}

As we have already pointed out, we are most interested in the limit
of large $\theta$ when the cusp anomalous dimension reproduces the
anomalous dimensions of twist-two conformal operators with large
spin. As $\theta\to \infty$ one approaches the light-cone.
Korchemsky and Marchesini~\cite{KM93} demonstrated how to calculate
the cusp anomalous dimension directly from the light-cone Wilson
loop:%
\footnote{An extra factor of 1/2 in this formula is due to our
regularization prescription.}
\begin{equation}
a \frac{\d}{\d a}
\ln W \left(\Gamma_{\rm l.c.}\right) = \frac 14 f(\lambda) \ln \frac Ta \,,
\label{KM}
\end{equation}
where $f(\lambda)$ is the same as in \eq{largetheta}.

We can obtain the light-cone ladder equation either directly by
summing the light-cone ladder diagrams or taking the
$\theta\to\infty$ limit of the expressions in Eqs.~\rf{tladder1} and
\rf{tladder2}. We find
\begin{equation}
\G_\alpha\left(S,T;a,b\right)=1-\bl \int_a^S ds \int_b^{ T} dt\,
\frac{\G_\alpha\left(S,T;s,t\right)}{\alpha s^2 + st} \,,
\label{nonladder1}
\end{equation}
where
\be
\bl=\frac\lambda{8\pi^2}
\label{bl}
\ee
and we have redefined $T\to T/2$ and introduced
\be
\alpha=\frac{u^2}{uv}=\pm 1
\ee
(remember that $v^2=0$ for the light-cone direction). If we
alternatively pick up the last (farthest from the cusp) rung of the
ladder,  we get
\begin{equation}
\G_\alpha\left(S,T;a,b\right)=1-\bl \int_a^S ds \int_b^{ T} dt\,
\frac{\G_\alpha\left(s,t;a,b\right)}{\alpha s^2 + st} \,.
\label{nonladder2}
\end{equation}
These are of the type of Eqs.~(\ref{tladder1}) and (\ref{tladder2}).

Differentiating Eq.~(\ref{nonladder2}) we obtain
\begin{equation}
S\frac{\partial}{\partial S}\,T\frac{\partial}{\partial T}\,
\G_\alpha\left(S,T;a,b\right)=
-\frac{\bl}{1+\alpha S/T}~
\G_\alpha\left(S,T;a,b\right).
\label{edifladder2}
\end{equation}
The differentiation of Eq.~(\ref{nonladder1}) analogously gives
\begin{equation}
a\frac{\partial}{\partial a}\,b\frac{\partial}{\partial b}\,
\G_\alpha\left(S,T;a,b\right)=
-\frac{\bl}{1+\alpha a/b}~
\G_\alpha\left(S,T;a,b\right),
\label{edifladder1}
\end{equation}
while \eq{ladexpo} is substituted by
\begin{equation}
F(S,T;a,b)= \bl \int_a^S ds \int_b^T dt \,
\frac{1}{s(\alpha s+t)} +\int_a^S ds \int_b^T dt \,
\frac{\partial F(s,t;a,b)}{\partial s}
\frac{\partial F(s,t;a,b)}{\partial t} \,.
\label{ladexpolc}
\end{equation}

The differential equations \rf{edifladder2} and  \rf{edifladder1}
should be supplemented with the boundary conditions
\begin{equation}
\G(a,T;a,b)=\G(S,b;a,b)=1\,.
\label{bc}
\end{equation}
These boundary conditions follow from the integral
equation~\rf{nonladder1} (or \rf{nonladder2}).

Analogously to \eq{tF1} we obtain
\begin{eqnarray}
F_1(S,T;a,b)&=&\bl \left( L_2(-\frac T{\alpha S}) -L_2(-\frac T{\alpha a})
-L_2(-\frac b{\alpha S})+L_2(-\frac b{\alpha a})  \right).
\label{F1}
\end{eqnarray}
Using Eq.~(\ref{Eul}), we can rewrite (\ref{F1}) in the equivalent form
\begin{equation}
F_1(S,T;a,b)= \bl \left( \ln \frac Tb \ln \frac Sa -L_2(-\alpha \frac ST)
+ L_2(-\alpha \frac aT)+L_2(-\alpha \frac Sb)-L_2(-\alpha \frac ab)
\right).
\label{F1al}
\end{equation}
This form is convenient to reproduce the order $\bl$ of
the $\alpha\rightarrow0$ limit, when the exact
$G$ is given by the Bessel function
\begin{equation}
\G_0= J_0(2\sqrt{\bl\ln \frac Sa \ln \frac Tb})\,.
\label{J0}
\end{equation}
The latter
formula for the $\alpha=0$ limit can be easily obtained by iterations
of the ladder equation \rf{nonladder1} (or \rf{nonladder2}).
Alternatively, the form (\ref{F1}) is convenient to find
the $\alpha\rightarrow\infty$ limit.

We can also rewrite (\ref{F1}) as
\begin{equation}
F_1(S,T;a,b)= \bl \left( L_2(-\frac T{\alpha S}) +L_2(-\frac {\alpha a}T)
+\frac12 \ln^2\frac T{\alpha a}+\frac{\pi^2}6
-L_2(-\frac b{\alpha S})+L_2(-\frac b{\alpha a})  \right).
\label{F1equi}
\end{equation}
If $S\rightarrow\infty$, we find from (\ref{F1equi})
\begin{equation}
F_1(S=\infty,T;a,b)=\bl \left( \frac12 \ln^2\frac T{\alpha a}+\frac{\pi^2}6
+L_2(-\frac b{\alpha a}) +L_2(-\frac {\alpha a}T) \right).
\label{F1fin}
\end{equation}
Remember that $L_2(0)=0$ (associated with $b=0$),
$L_2(-1)=-\pi^2/12$ from Eq.~(\ref{Eul})
(associated with $\alpha a=b=\epsilon$)
and $L_2(1)=\pi^2/6$ from Eq.~(\ref{Eul})
(associated with $\alpha=-1$, $ a=b=\varepsilon$).

For $S\sim T \gg a,b$, we have from (\ref{F1equi})
\begin{equation}
F_1(S,T;a,b)\stackrel{S,T\gg a,b}=
\bl \left( L_2(-\frac T{\alpha S}) +
\frac12 \ln^2\frac T{\alpha a}+\frac{\pi^2}6
+L_2(-\frac b{\alpha a})  \right)
\end{equation}
and
\begin{equation}
F_1(S=\infty,T;a,b)
\stackrel{T\gg a,b}=\bl \left( \frac12 \ln^2\frac T{\alpha a}+\frac{\pi^2}6
+L_2(-\frac b{\alpha a})  \right).
\label{F1finas}
\end{equation}

The order $\bl^2$ can be obtained by inserting (\ref{F1}) into the
second term on the right-hand side of Eq.~(\ref{ladexpo}). This
gives
\begin{equation}
F_2(S,T;a,b) = \bl^2 \int_a^S ds_1 \int_b^T dt_2
\int_b^{t_2} \frac{dt_1}{s_1(\alpha s_1+t_1)}
\int_a^{s_1} \frac{ds_2}{s_2(\alpha s_2+t_2)}
\label{crossedladders}
\end{equation}
which is nothing but the light-cone diagram with crossed ladders.
Integrating over $s_2$ and $t_1$, we find
\begin{equation}
F_2(S,T;a,b) = \bl^2 \int_a^S
\frac {ds}s \int_b^T \frac{dt}t \ln\frac {\alpha+t/s}{\alpha+b/s}
\ln \frac{\alpha+t/a}{\alpha+t/s} \,.
\label{iF2Int}
\end{equation}

Also we obtain
\begin{eqnarray}
\lefteqn{\left(a\frac {d}{da} \right)^2 F_2(S=\infty,T;a,b)} \non &&
\stackrel{T\gg a,b}= \bl^2 \left[
\frac 12 \ln^2 \frac T{\alpha a}+\frac {\pi^2}6 +L_2(-\frac b{\alpha a})
-\ln \left( 1+ \frac b{\alpha a} \right)
\ln\frac{( \alpha a+ b )}{T} \right].
\end{eqnarray}
With logarithmic accuracy this yields
\begin{equation}
F_2(S=\infty,T;a,b\sim a)\stackrel{T\gg a}
= \bl^2 \left( \frac1{24}\ln^4 \frac T{\alpha a} +
\frac{\pi^2}{12}\ln^2 \frac T{\alpha a} \right).
\label{F2fin}
\end{equation}

The second term on the right-hand side of Eq.~(\ref{F2fin}) is
of the same type as in (\ref{F1fin}) which contributes to the cusp
anomalous dimension. For $\alpha=1$ it agrees with \eq{lc}.

For $\alpha=-1$ we get from Eq.~(\ref{F2fin})
\begin{equation}
{\rm Re}\;F_2(S=\infty,T;a,b\sim a)\stackrel{T\gg a}=
\bl^2 \left( \frac1{24}\ln^4 \frac T{a} -
\frac{\pi^2}{6}\ln^2 \frac T{a} \right)
\label{al=-1F2fin}
\end{equation}
again with logarithmic accuracy.
This is to be compared with the evaluation of the integral on the
right-hand side of Eq.~(\ref{iF2Int}) for $S=T$, $b=a$ and
$\alpha=-1$ which results in the exact formula
\begin{equation}
{\rm Re}\;F_2(S=T;a=b)\stackrel{\alpha=-1}
= \bl^2 \left( \frac1{24}\ln^4 \frac Ta -
\frac{\pi^2}{6}\ln^2 \frac Ta \right)
\label{al=-1F2fina}
\end{equation}
for all values of $T$.
The fact that the $\ln^2(T/a)$ term is the same as in Eq.~(\ref{al=-1F2fin})
confirms the expectation that the cusp anomalous dimension can
be extracted from the $S\sim T\gg a\sim b$ limit, which is based on the fact
that Log's of $S/a$ never appear in perturbation theory.

The appearance of the $\ln^4(T/a)$ term in Eqs.~(\ref{F2fin}) and
(\ref{al=-1F2fin}) implies no exponentiation for the ladder
diagrams. It is already known from Sect.~\ref{s:3}
that the exponentiation occurs only when
the ladders are summed up with the anomalous term observed
in Sect.~\ref{s:2}.

\subsection{Exact solution for light-cone ladders}

To satisfy Eqs.~(\ref{edifladder2}) and (\ref{edifladder1}),
we substitute the ansatz
\begin{equation}
\G_\alpha\left(S,T;a,b\right) =\oint\limits_C \frac{d\omega}{2\pi i \omega}
\left(\frac S{a} \right)^{\sqrt{\bl}\omega }
\left(\frac T{b} \right)^{-\sqrt{\bl}\omega^{-1} }
\,F\left(-\omega, \alpha \frac ab \right)
F\left(\omega, \alpha \frac ST \right),
\label{ansatz}
\end{equation}
where $C$ is a contour in the complex $\omega$-plane.
This ansatz is motivated by the integral representation of the Bessel
function~\rf{J0} at $\alpha=0$.
The substitution into Eq.~(\ref{edifladder2}) reduces it to the
hypergeometric equation ($\xi =\alpha S/T$)
\begin{equation}
\xi(1+\xi) F''_{\xi\xi}+ [1+\sqrt{\bl}(\omega+\omega^{-1})](1+\xi)F'_\xi
+\bl F=0
\label{hg}
\end{equation}
whose solution is given by hypergeometric functions.
The same is true for Eq.~(\ref{edifladder1}) when $\omega$ is substituted
by $-\omega$ and $\xi= \alpha a/b$.

We have shown that the following combination of solutions
satisfies the boundary conditions~\rf{bc}:
\begin{eqnarray}
\lefteqn{G_\alpha (S,T;a,b)=\oint\limits_{C^r} \frac{d\omega}{2\pi i\omega}~
{}_2 F_1\left(-\sqrt{\bl}\omega,
-\sqrt{\bl}\omega^{-1};
1-\sqrt{\bl}(\omega+\omega^{-1});-\alpha\frac ab \right)}
\nonumber \\* && \hspace*{1cm}
\times\left( \frac Sa \right)^{\sqrt{\bl}\omega }
\left(\frac Tb \right)^{-\sqrt{\bl}\omega^{-1} }
{}_2 F_1 \left( \sqrt{\bl}\omega ,\sqrt{\bl}\omega^{-1};
1+\sqrt{\bl}(\omega +\omega^{-1});
-\alpha \frac ST  \right) \nonumber \\
&&+ \int_{|_ {n_{\rm min}-0}} \frac {d s}{2\pi i}
\frac{\Gamma(-s)}{\Gamma(s)}
\frac{1}{\sqrt{\bl}(\omega_+^R-\omega_-^R)}
\frac{\Gamma(\sqrt{\bl}\omega_+^R)\Gamma(1+\sqrt{\bl}\omega_+^R)}
{\Gamma(\sqrt{\bl}\omega_+^L)\Gamma(1+\sqrt{\bl}\omega_+^L)}
\nonumber \\* && \hspace{1cm} \times \left(\alpha\frac ab\right)^s
\left[\left(\frac Sa\right)^{\sqrt{\bl}\omega_+^R}\,
\left(\frac Tb\right)^{-\sqrt{\bl}\omega_-^R} +
 \left(\frac Sa\right)^{\sqrt{\bl}\omega_-^R}\,
\left(\frac Tb \right)^{-\sqrt{\bl}\omega_+^R}  \right]
\nonumber \\* && \hspace*{1cm} \times
{}_2 F_1\left(\sqrt{\bl}\omega_+^R,
\sqrt{\bl}\omega_-^R; s+1;-\alpha\frac ab \right)
{}_2 F_1 \left( \sqrt{\bl}\omega_+^R ,\sqrt{\bl}\omega_-^R;
s+1; -\alpha \frac ST  \right) \nonumber \\*
&&
\label{aaremarkable}
\end{eqnarray}
with
\begin{eqnarray}
\omega^R_\pm(s)  &= &\frac s{2\sqrt{\bl}} \pm \sqrt{\frac{s^2}{4\bl}-1}
\,,
\nonumber \\
\omega^L_\pm (s) &= & -\omega_\mp^R=
-\frac s{2\sqrt{\bl}} \pm \sqrt{\frac{s^2}{4\bl}-1}
\,.
\label{sppoles}
\end{eqnarray}
The contour integral in the first term on the right-hand side of
\eq{aaremarkable} runs over a circle of arbitrary radius $r$
($|\omega|=r$), while the second contour integral runs parallel to
imaginary axis along the line
\begin{equation}
s=n_{\rm min}-0 +ip\qquad(-\infty<p<+\infty) \,,
\end{equation}
where
\begin{equation}
n_{\rm min}=\left[\sqrt{\bl}(r+1/r)\right]+1
\label{nmin}
\end{equation}
and $[\cdots]$ denotes the integer part.

Each of the two terms on the right-hand side of
Eq.~(\ref{aaremarkable}) satisfies the linear differential equations
(\ref{edifladder2}) and (\ref{edifladder1}). They are added in the
way to satisfy the boundary conditions~\rf{bc}.

To show this,
it is crucial that the poles at $\omega=\omega_\pm^R(n)$
and $\omega=\omega_\pm^L(n)$ of the integrand in the first term
are canceled by the poles at $s=n$ of the integrand in the second term.
The position of the real part of the contour of integration over $s$
is such to match the number of poles. If $r<1$ the poles at $\omega_-^R(n)$
and $\omega_+^L(n)$ lie inside the circle of integration for
$\sqrt{\bl}<(r+r^{-1})^{-1}$.
This corresponds to  $n_{\rm min}=1$ in \eq{nmin} so that all poles
of the integrand in the second term are to the right of the
integration contour over $s$. The residues are chosen to be the same.
At $\sqrt{\bl}=(r+r^{-1})^{-1}$ the poles at $\omega_-^R(1)$
and $\omega_+^L(1)$ crosses the circle of integration and correspondingly
the contour of
integration over $s$ jumps toward ${\rm Re}\,s=2$ because now $n_{\rm min}=2$.
With increasing $\sqrt{\bl}$
the poles at $\omega_-^R(n)$
and $\omega_+^L(n)$ with $n\geq n_{\rm min}$ of the integrand in the first
term, which lie inside the circle of integration,
are canceled by the poles at $s=n\geq n_{\rm min}$ of the integrand
in the second term. A useful formula which provides the cancellation
is
\begin{equation}
{}_2F_1 (A,B;C;z)\stackrel{C\rightarrow -n+1}
\longrightarrow \Gamma (C) \frac{\Gamma(A+n)\Gamma(B+n)}
{\Gamma(A)\Gamma(B)\,n!}\,z^n{}_2F_1 (A+n,B+n;n+1;z)
\label{4}
\end{equation}
as $C\rightarrow -n+1$.

To demonstrate how the boundary conditions~\rf{bc} are satisfied by the
solution~\rf{aaremarkable}, we choose the integration contour in
the first term to be a circle of the radius which is
either very small for $T=b$ or very large for $S=a$.
Then the form of the integrand is such that
the residue at the pole, respectively, at $\omega=0$ or $\omega=\infty$
equals 1 which proves that the boundary conditions is satisfied.

The numerical value of $\G$ given by \eq{aaremarkable}
can be computed for a certain range of the parameters
$S$, $T$, $b$ and $\alpha$ using the Mathematica
program in Appendix~A.

\subsection{Exact solution for light-cone ladders (continued)}

A great simplification occurs in \eq{aaremarkable} for $\alpha=-1$,
$S=T$ and $b=a$, when the hypergeometric functions reduce
to gamma functions:
\be
{}_2F_1 \left(A,B;1+A+B;1  \right)= \frac{\Gamma\left( 1+A+B\right)}
{\Gamma\left( 1+A\right)\Gamma\left( 1+B\right)}\,.
\ee
We then obtain
\begin{eqnarray}
\lefteqn{\hspace*{-.4cm} G_{\alpha=-1} (T,T;a,a)=
\oint\limits_{C^r} \frac{d\omega}{2i \pi^2 \sqrt{\bl}\omega}~
 \left(\frac Ta\right)^{\sqrt{\bl}(\omega-\omega^{-1})}\,
\frac{\sin{(\pi \sqrt{\bl}\omega)}\sin{(\pi \sqrt{\bl}\omega^{-1})}}
{\sin{(\pi \sqrt{\bl}(\omega+\omega^{-1}))}}\,(\omega+\omega^{-1})}
\nonumber \\*
&&+ \int_{|_ {n_{\rm min}-0}} \frac {d s}{2i\bl \pi^2}
\frac{s(-1)^s}{\sin(\pi s)}
\frac{\sin^2(\pi\sqrt{\bl}\omega_-^R)}
{\sqrt{\bl}(\omega_+^R-\omega_-^R)}
\left[\left(\frac Ta\right)^{\sqrt{\bl}(\omega_+^R-\omega_-^R)} +
 \left(\frac Ta\right)^{\sqrt{\bl}(\omega_-^R-\omega_+^R)}  \right].
 \nonumber \\*
&&
\label{abremarkable}
\end{eqnarray}

We can now make a change of the integration variable
in the first contour integral from $\omega$ to
\begin{equation}
s=\sqrt{\bl}(\omega+\omega^{-1})
\label{svso}
\end{equation}
so that
\begin{equation}
\frac{d\omega}{\omega} =\frac {ds}{\sqrt{s^2-4\bl}}
\label{dsdo}
\end{equation}
and rewrite \eq{abremarkable} as
\begin{eqnarray}
\lefteqn{\G_{\alpha=-1} (T,T;a,a)}\non&=&
\int\limits_{-2\sqrt{\bl}}^{+2\sqrt{\bl}}
\frac{d s}{2 \pi^2 {\bl}}\,
\frac{s}{\sqrt{\bl-s^2/4}}\,
 \left(\frac Ta\right)^{\sqrt{\bl}(\omega_+^R-\omega_-^{R})}\,
\frac{\sin{(\pi \sqrt{\bl}\omega_+^R)}
\sin{(\pi \sqrt{\bl}\omega_-^{R})}}
{\sin{(\pi s)}}
\nonumber \\*
&&+ \int_{|_ {n_{\rm min}-0}} \frac {d s}{2i\bl \pi^2}
\frac{s(-1)^s}{\sin(\pi s)}
\frac{\sin^2(\pi\sqrt{\bl}\omega_-^R)}
{\sqrt{\bl}(\omega_+^R-\omega_-^R)}
\left[\left(\frac Ta\right)^{\sqrt{\bl}(\omega_+^R-\omega_-^R)} +
 \left(\frac Ta\right)^{\sqrt{\bl}(\omega_-^R-\omega_+^R)}  \right].
 \nonumber \\*
&&
\label{newaaremarkable}
\end{eqnarray}
The integrand of the first term on the right-hand side may have
poles for $\bl>1/4$ at $s=\pm n$.  Then it should be understood as
the principal value integral.

For $\sqrt{\bl}<1/2$ we can first substitute $s\rightarrow -s$ in
the second term in the square brackets in the last line of
Eq.~(\ref{newaaremarkable}) to get the integral over ${\rm
Re}\,s=-1$ and then to deform the contour of the integration from
${\rm Re}\,s=\pm1$ to obtain the same contour as in the first
integral on the right-hand side.

We then rewrite \eq{newaaremarkable} as
\begin{eqnarray}
\lefteqn{\G_{\alpha=-1} (T,T;a,a)}\non&=&
\int\limits_{-2\sqrt{\bl}}^{+2\sqrt{\bl}}
\frac{d s}{2 \pi^2 {\bl}}\,
\frac{s \sin{(\pi s)}}{\sqrt{4\bl-s^2}}\,
\cos{\left( \sqrt{4\bl-s^2}(\ln \frac Ta-i\pi)\right)}
\,
\sin{\left(\pi \sqrt{4\bl-s^2}\right)}.
\nonumber \\*&&
\label{nnewaaremarkable}
\end{eqnarray}
Consulting with the table of integrals%
\footnote{I.S. Gradshteyn and I.M. Ryzhik, Table of integrals, series and
products, Eq.~3.876.7 on p.~473.},
we finally express the integral in \eq{nnewaaremarkable}
via the Bessel function
\begin{equation}
\G_{\alpha=-1} (a e^{\tau},a e^{\tau} ;a,a)=
\frac{1}{\sqrt{\bl \tau (\tau - 2 \pi i)}} \,
J_1 \left(2 \sqrt{\bl \tau (\tau - 2 \pi i)}   \right)
\label{GfinJ}
\end{equation}
with
\be
\ln \frac Ta=\tau, \qquad
\ln \left(-\frac Ta\right)= \tau - i\pi
\ee
and $\bl$ given by \eq{bl}.
Note this is $J_1$ rather than $I_1$ as in Ref.~\cite{ESZ00}
because of Minkowski space.

Taking according to \eq{FG} $-$Log of the expansion of~(\ref{GfinJ}) in $\bl$,
we get
\begin{eqnarray}
F_{\alpha=-1} (a e^{\tau},a e^{\tau} ;a,a)&=&
\frac12\Big(\bl \tau (\tau-2\pi i)\Big)+
\frac1{24}\Big(\bl \tau (\tau-2\pi i)\Big)^2+
\frac1{144}\Big(\bl \tau (\tau-2\pi i)\Big)^3 \nonumber \\*
&&+
\frac1{720}\Big(\bl \tau (\tau-2\pi i)\Big)^4+
{\cal O}\left( \bl^5 \right).
\label{Ffinpt}
\end{eqnarray}
The order $\bl^2$ is in a perfect agreement with Eq.~(\ref{al=-1F2fin}).
We see that only the first two ladders have a $\tau^2$ term
and, therefore, contribute to the cusp anomalous dimension.

\subsection{Asymptotic behavior}

It is easy to write down the asymptote of the solution~\rf{GfinJ}
at large $\tau=\ln\frac Ta$:
\be
\G_{\alpha=-1} (a e^{\tau},a e^{\tau} ;a,a)\stackrel{\tau\gg1}\approx
\cos{\left(2\sqrt{\bl} \tau- \frac{3\pi}4 \right)}
\frac1{\sqrt{\pi} (\sqrt{\bl} \tau)^{3/2}}\,.
\label{asympt1}
\ee
One may wonder how this asymptote can be extended to the case of
$S\neq T$, in particular $S \gg T$.

Rewriting the differential ladder equation~\rf{edifladder2} via
the variables
\be
x=\ln \frac Sa - \ln \frac Ta \,,\qquad y =\ln \frac Sa + \ln \frac Ta \,,
\ee
we obtain
\be
\left( \frac {\partial^2}{\partial x^2}
-\frac {\partial^2}{\partial y^2}\right) \G (x,y)=
\frac{\bl}{1-  e^{ x}} \;\G (x,y)
\label{eadlad}
\ee
for $\alpha=-1$ and $b=a$.

The existence of the stationary-phase points of both integrals on
the right-hand side of \eq{newaaremarkable} at $s=0$ valid for $S=T$
suggests to satisfy \eq{eadlad} at large $\bl$ by introducing a
phase in the argument of the cosine in \eq{asympt1}: \be
\cos{\left(2\sqrt{\bl} \tau- \frac{3\pi}4 \right)} \Longrightarrow
\cos{\left(\sqrt{\bl} y -\sqrt{\bl}\varphi(x)- \frac{3\pi}4
\right)}\,. \label{phase} \ee Then \eq{eadlad} is satisfied by the
cosine for large $\bl$ if \be \varphi(x) = 2\, {\rm arccosh}\,
e^{x/2} \label{vphi} \ee which obeys $\varphi(0)=0$ as it should. In
a more rigorous treatment the cosine should of course be multiplied
by a decreasing prefactor.

As we expected from the analysis of perturbation theory,
where each term remains finite as $S\to\infty$,
the arguments of the cosine on both sides of \eq{phase}
should remain the same with logarithmic accuracy as $S\to\infty$.
This is indeed the case for the solution~\rf{vphi} which
behaves at large $x$ as
\be
\varphi(x) = x-2 \ln 2 +{\cal O} (x^{-1}) \,,
\ee
so that we find
\be
\cos{\left(\sqrt{\bl} y -\sqrt{\bl}\varphi(x)- \frac{3\pi}4 \right)}
\stackrel{S\to\infty}=
\cos{\left(2 \sqrt{\bl} \tau +2\sqrt{\bl}\ln 2- \frac{3\pi}4 \right)}\,.
\ee

This type of the asymptotic behavior is not of the type given by \eq{KM}
and leads us to the conclusion that
the ladder diagrams cannot reproduce the
$\sqrt{\lambda}$-behavior of the cusp anomalous dimension for large
$\lambda$.

\newsection{Discussion\label{s:d}}

The main conclusion of this paper is that the cusped Wilson loop
possesses a number of remarkable dynamical properties which make it
a very interesting object for studying the string/gauge
correspondence. On one hand its dynamics in ${\cal N}=4$ SYM is more
complicated than that of the solvable cases of the straight line or
circular loop and it therefore is a more powerful probe of the gauge
theory. On the other hand it has certain simple features which could
be accessible to analytic computations and which could be universal,
in the sense that they are shared by non-supersymmetric Yang-Mills
theory or even QCD.

An example of the latter is the appearance of the anomaly term in
the two-loop calculations of this paper. Unlike the case of smooth
loops, the cancellation of the divergent parts of Feynman diagrams
with internal vertices is not complete. What remains has a nice
simple structure of an ``anomalous surface term'' where some legs of
the diagrams with internal vertices are frozen onto the cusp. This
is reminiscent of the anomaly explanation of the simple structure of
the circular loop and its subsequent relation to a one-matrix model
presented in Ref.~\cite{DG00}.  In the case of the cusp, the
``degrees of freedom'' which give rise to the divergent part of the
expectation value seem to reside at the location of the cusp. We
expect this to persist in higher orders of perturbation theory.  The
fact that the sum of all ladders does not seem to contribute to the
leading term in the cusp anomalous dimension means that the entire
contribution, if it is indeed there, comes from diagrams with
internal vertices. We expect that such diagrams all have legs frozen
at the location of the cusp like the diagram
depicted in Fig.~3(a).  It would be interesting to try to
characterize these diagrams and understand the generic contribution.

Analogously, the loop equation reveals a number of interesting
properties which are specific to the cusped loop. Here, we have
checked that the appearance of the cusp anomalous dimension
in the supersymmetric loop equation observed
to one-loop in Ref.~\cite{DGO99}, actually persists at two
loop order. This involves some non-trivial and rather surprising
identities for integrals that we have detailed in Sect.~4.

Finally, the sum of the ladder diagrams can be found exactly and for
certain values of parameters reduces to the Bessel function. In
fact, the Bessel function is very similar to the one that is thought
to be the exact expression for the circular loop~\cite{ESZ00}.

It is not yet clear which of these features are common with usual
Yang-Mills theory, but some of them certainly are. One of such
quantity is the universal part of the cusp anomalous dimension in
pure QCD at two loops, which does not depend on the regularization
prescription. It may support the expectation put forward in
Ref.~\cite{GKP02} that the universal part of the anomalous
dimensions of twist-two operators with large spin $J$ in pure
Yang-Mills theory and ${\cal N}=4$ SYM are in fact identical.

In order to clarify this assertion, we compare the anomalous
dimension of the cusped SYM Wilson loop, calculated in this paper,
with the analogous calculation of the one for the properly
regularized non-supersymmetric Wilson loop of only Yang-Mills field
in Yang-Mills theory with adjoint matter. 
While the
fermionic contribution has been known for a while~\cite{KR87}, the
contribution from scalars has been calculated relatively
recently~\cite{BGK03}.
The result is given at
large $\theta$ by \be \gamma_{\rm cusp}
=\frac{\theta}2 \left[ \frac{\lambda}{2\pi^2}
+\frac{\lambda^2}{24\pi^4} \left(\frac{16}3 -\frac{\pi^2}{4}-\frac
56 n_f -\frac 13 n_s \right) \right] +{\cal
O}\left(\lambda^3\right), \label{BGK} \ee 
where $n_f$ is the number
fermionic species and $n_s$ is the number of scalars (which are
present only in the action but not in the definition of the Wilson
loop as is already said). 
The pure Yang-Mills contribution
(associated with $n_f=n_s=0$) is regularization-dependent at order
$\lambda^2$ and has a universal part $\propto \lambda^2/\pi^2$ as
well as the regularization-dependent part $\propto \lambda^2/\pi^4$.
Here, the latter, regularization-dependent part is written for
regularization via dimensional reduction (the DR scheme).

For the ${\cal N}=4$ SYM we substitute in \eq{BGK} $n_f=4$ and
$n_s=6$ after which the non-universal part vanishes and we reproduce
\eq{cusp2}. This means that the only effect of scalars (as well as
of fermions) is their contribution to the renormalization of
the coupling constant,
while they decouple from the supersymmetric light-cone Wilson loop
because $\dot x^2=0$ at the light-cone.

For the latter reason it would be interesting to investigate loop 
equation \rf{cle} 
for supersymmetric cusped Wilson loops
in ${\cal N}=4$ SYM. A nice property of this
equation is that the contribution of scalars to the right-hand side
vanishes at the light-cone. The cusped loop equation, which sums up
all relevant planar diagrams, can be also useful for calculations of
the cusp anomalous dimension. In fact the structure of the
right-hand side of \eq{cle} is such that it involves only one cusped
Wilson loop while another $W$ is rather a Bethe-Salpeter kernel of
the type calculated in this paper. These issues deserve further
investigation.

\subsection*{Acknowledgments}

Y.M. is indebted to J.~Ambj\o rn, A.~Gorsky, H.~Kawai, L.~Lipatov,
J.~Maldacena, A.~Tsey\-tlin, and K.~Zarembo for useful discussions.
The work of Y.M. was partially supported by the Federal Program of
the Russian Ministry for Industry, Science and Technology
No 40.052.1.1.1112.
The work of Y.M. and P.O. was supported in part by the grant
{\mbox INTAS~03--51--5460}.
The work of Y.M. and G.S. was supported in part
by the grant NATO~CLG--5941.
G.S. acknowledges financial support of NSERC of Canada.

\setcounter{section}{0}
\appendix{Program for computing (\ref{aaremarkable})}

\begin{verbatim}
(* the value of x = Sqrt[beta] *)
x =.
(* the values of S, T, b, al = alpha *)
S = 100
T = 100
a = 1
b = a
al = 1
R = .8
(* residues are summed up from n = Real[s] + 1 *)
s[x_, p_] := IntegerPart[x(R + 1/R)] + .9999 + I p
ORp[ss_] := ss/2 + Sqrt[ss^2/4 - x^2]
ORm[ss_] := ss/2 - Sqrt[ss^2/4 - x^2]
OLp[ss_] := -ss/2 + Sqrt[ss^2/4 - x^2]
OLm[ss_] := -ss/2 - Sqrt[ss^2/4 - x^2]
(* enumeration of the contour integral *)
Int[x_, T_] := NIntegrate[ (S/a)^(x R Exp[I phi])
   (T/b)^(-x R^(-1)Exp[-I phi])
   Hypergeometric2F1[-x R Exp[I phi], -x R^(-1)Exp[-I phi],
      1 - x R Exp[I phi] - x R^(-1)Exp[-I phi], -al a/b]
   Hypergeometric2F1[x R Exp[I phi], x R^(-1)Exp[-I phi],
      1 + x R Exp[I phi] + x R^(-1)Exp[-I phi], -al S/T]/(2 Pi),
    {phi, 0, 2 Pi}, MaxRecursion -> 16]
(* the sum over residues inside the circle *)
Res[x_, T_] :=
  NIntegrate[(Gamma[-s[x, p]]/
          Gamma[s[x, p]])(ORp[s[x, p]] - ORm[s[x, p]])^(-1)( al a/b)^
        s[x, p](Gamma[ORp[s[x, p]]]Gamma[ORp[s[x, p]] +
                1]/(Gamma[OLp[s[x, p]]]Gamma[OLp[s[x, p]] + 1]))
                  ((S/a)^(ORp[s[x, p]])(T/b)^(-ORm[s[x, p]])
                     + (S/a)^(ORm[s[x, p]])(T/b)^(-ORp[s[x, p]]))
   Hypergeometric2F1[-OLp[s[x, p]], -OLm[s[x, p]], 1 + s[x, p], -al a/b]
   Hypergeometric2F1[ORp[s[x, p]], ORm[s[x, p]],
          1 + s[x, p], -al S/T]/(2 Pi), {p, -Infinity, +Infinity},
    MaxRecursion -> 16]
G[x_, T_] := Int[x, T] + Res[x, T]
    (* G[x, T] *)
Plot[G[x, T], {x, 0, 3.}]

\end{verbatim}


\end{document}